\documentclass{jfm}

\usepackage{graphicx}
\usepackage{epstopdf,epsfig}
\usepackage{newtxtext}
\usepackage{newtxmath}
\usepackage{natbib}
\usepackage{hyperref}
\usepackage{xcolor} 
\hypersetup{
    colorlinks = true,
    urlcolor   = blue,
    citecolor  = black,
}
\usepackage{caption}
\newcommand{\RomanNumeralCaps}[1]

\title{Large-to-small scale frequency modulation analysis in wall-bounded turbulence via visibility networks}

\author{Giovanni Iacobello\aff{1}
  \corresp{\email{giovanni.iacobello@polito.it}},
  Luca Ridolfi\aff{2},
 \and Stefania Scarsoglio\aff{1}}

\affiliation{\aff{1}Department of Mechanical and Aerospace Engineering, Politecnico di Torino, Turin, Italy
\aff{2}Department of Environmental, Land and Infrastructure Engineering, Politecnico di Torino, Turin, Italy}

\begin{document}
\maketitle

\begin{abstract}

	Scale interaction is studied in wall-bounded turbulence by focusing on the frequency modulation (FM) mechanism of large scales on small scale velocity fluctuations. Differently from amplitude modulation analysis, frequency modulation has been less investigated also due to the difficulty to develop robust tools for broadband signals. To face this issue, the natural visibility graph approach is proposed in this work to map the full velocity signals into complex networks. We show that the network degree centrality is able to capture the signal structure at local scales directly from the full signal, thereby quantifying FM. Velocity signals from numerically-simulated turbulent channel flows and an experimental turbulent boundary layer are investigated at different Reynolds numbers. A correction of Taylor's hypothesis for time-series is proposed to overcome the overprediction of near-wall frequency modulation obtained when local mean velocity is used as the convective velocity. Results provide network-based evidences of the large-to-small FM features for all the three velocity components in the near-wall region, with a reversal mechanism emerging far from the wall. Additionally, scaling arguments in the view of the quasi-steady quasi-homogeneous hypothesis are discussed, and a delay-time between large and small scales very close to the near-wall cycle characteristic time is detected. Results show that the visibility graph is a parameter-free tool that turns out to be effective and robust to detect FM in different configurations of wall-bounded turbulent flows. Based on present findings, the visibility network-based approach can represent a reliable tool to systematically investigate scale interaction mechanisms in wall-bounded turbulence.

\end{abstract}


\section{Introduction}\label{sec:Introd}
		
		The characterization and modelling of wall-bounded turbulent flows is of paramount importance in physics and engineering~\citep{marusic2010predictive}. Organized motions, in particular, play a crucial role in wall-bounded turbulence analysis, since they are associated to high energy levels and are directly involved in transport processes, making them preferential targets for flow control strategies~\citep{jimenez2018coherent}. Coherent streaks are recognized as the dominant flow structures very close to the wall, and are characterized by a distinctive (inner) peak in the spectrogram of the streamwise velocity fluctuations, $u$, within the buffer layer~\citep{jimenez2018coherent}. The investigation of high Reynolds number experiments and simulations also revealed the formation of large scale motions (LSMs) and very large scale motions (VLSMs) residing in the log-region~\citep{smits2011high}, whose presence is detected by the appearance of another (outer) peak in the (pre-multiplied) energy spectrogram of the streamwise velocity fluctuations~\citep{hutchins2007large, monty2009comparison, peruzzi2020scaling}. The wall-normal location in wall units (i.e., made dimensionless by the mean friction velocity, $U_\tau$, and the fluid kinematic viscosity, $\nu$), $y^+=yU_\tau/\nu$, of the inner peak is conventionally assumed to be fixed at $y^+=15$, while the position of the outer peak increases with the frictional Reynolds number, $\Rey_\tau$, as $y^+\approx 3.9\Rey_\tau^{1/2}$~\citep{mathis2009Large}. 
		
		\begin{figure}
	 		\centerline{\includegraphics{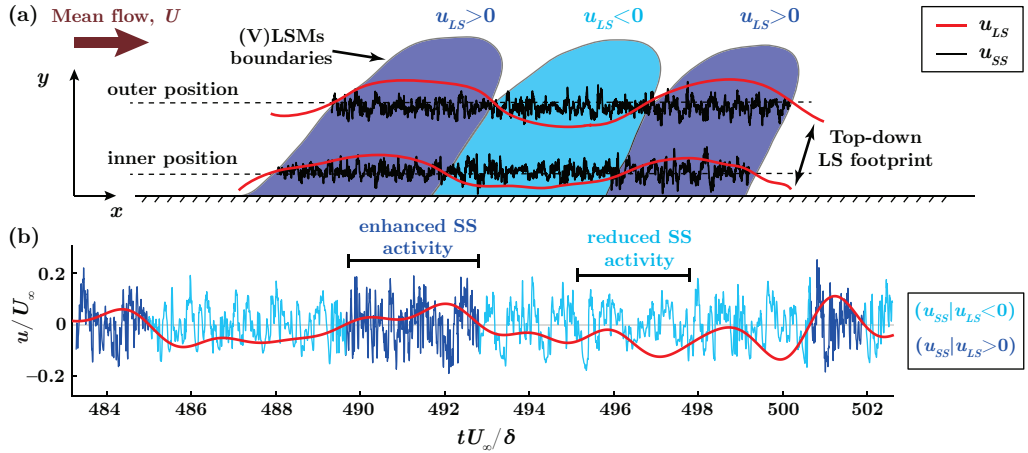}}
		  	\caption{(a) Schematic of a wall-bounded turbulent flows in the $(x$-$y)$ plane, showing three alternating LSM and VLSM structures of uniform large scale momentum, $u_{LS}\lessgtr 0$. Two pairs of time-series of $u_{LS}$ and $u_{SS}$ are also depicted as red and black lines, respectively, at two wall-normal locations (referred to as inner and outer position). (b) The small scale, $u_{SS}$, and the large scale, $u_{LS}$, of streamwise velocity fluctuations at $y^+\approx 10$ are shown as blue and red lines, respectively, where light and dark blue portions of $u_{SS}$ correspond to intervals of $u_{LS}<0$ and $u_{LS}>0$, respectively. The velocity series is extracted from an experimental turbulent boundary layer at $\Rey_\tau=14750$~\citep{MARUSIC2020} in the range $tU_\infty/\delta=483-503$, where $U_\infty$ and $\delta$ are the free stream velocity and boundary layer thickness, respectively. Two intervals of the signal (i.e., $490<t U_\infty /\delta <493$ and $495<t U_\infty /\delta <498$) in which small scales display enhanced or reduced activity are also indicated.}
			\label{fig:egserie_SSLS}
		\end{figure}

		Besides the effect of the Reynolds number, some differences emerged from the comparison of different canonical wall-bounded turbulent flows. While near-wall statistics (such as the mean velocity profile) agree well in channel, pipe and boundary layer flows, the features of large scales depend on the flow configuration~\citep{monty2007large, balakumar2007large, mathis2009comparison, monty2009comparison, chernyshenko2020extension}. In particular, spectral analyses of internal and external flows have revealed that very-large scales tend to be longer for channel and pipe flows than boundary layer flows, although they appear to be qualitatively similar ~\citep{balakumar2007large, monty2009comparison}. Such large scale differences are expected to grow for larger Reynolds numbers as energetic contributions coming from very-large scale motion increases with the Reynolds number.
		
		The investigation of higher Reynolds number data has progressively opened novel developments and questions about an interaction between small scale turbulence (whose spectral peak occurs in the wall proximity) and large scale motions (whose spectral peak resides far from the wall). First insights on scale interaction were initially reported by~\citet{brown1977large} and~\citet{bandyopadhyay1984coupling}, who observed a modulation mechanism of (near-wall) small scales by large turbulent scales. Later on,~\citet{hutchins2007large} provided further evidences of a top-down footprint and an amplitude modulation (AM) phenomenon by the large scale motions residing in the log-region on the near-wall (small scale) dynamics. Aiming to illustrate such inter-scale mechanism, \figurename~\ref{fig:egserie_SSLS}(a) shows a schematic of a wall-bounded turbulent flow in a streamwise-vertical plane, where uniform momentum regions due to LSM and VLSM (highlighted as dark and light blue structures) entail large scale fluctuations, $u_{LS}$ (see red lines). \figurename~\ref{fig:egserie_SSLS}(a) is drawn following the current picture of the kinematics of turbulent scales and their interaction in wall-bounded turbulence~\citep[e.g., see ][]{ganapat2012amplitude, baars2017reynolds}. The turbulent flow field can be decomposed as $u(x,y,z,t)=u_{LS}(x,y,z,t)+u_{SS}(x,y,z,t)$, where $u_{SS}$ are small scale fluctuations (see black signals), while $x,y,z$ are the streamwise, vertical and spanwise directions, respectively, and $t$ is time. \figurename~\ref{fig:egserie_SSLS}(a) also highlights the top-down footprint of large scales, being the two $u_{LS}$ signals (red lines) positively correlated with each other (eventually accounting for the inclination of large scales~\citep{marusic2007reynolds}). 
		\\A modulation of the amplitude of the small scales caused by the large scales implies that high or low values of $u_{LS}$ correspond to (on average) high or low values of $u_{SS}$. This mechanism can be observed in \figurename~\ref{fig:egserie_SSLS}(b), which shows a time interval of $u_{LS}$ and $u_{SS}$ at $y^+\approx 10$ in an experimental turbulent boundary layer. An increase of the local amplitude of the small scale signal (see dark blue intervals) is discernible during positive large scale velocity fluctuations, $u_{LS}>0$, and, \textit{vice versa}, a damping of small scale amplitudes (light blue intervals) during negative large scale velocity fluctuations, $u_{LS}<0$.

		\citet{mathis2009Large} quantified this amplitude modulation by correlating $u_{LS}$ with the large-scale-filtered envelope of $u_{SS}$ at different wall-normal coordinates. The authors evidenced an amplitude modulation (as shown in \figurename~\ref{fig:egserie_SSLS}(b)) only close to the wall (approximatively below the center of the log-region), while a reversed AM mechanism -- i.e., an $u_{SS}$ amplitude increase under $u_{LS}<0$ and an $u_{SS}$ amplitude decrease under $u_{LS}>0$ -- occurs far from the wall. Further studies on turbulent boundary layers have suggested that a modulation mechanism does actually take place only in the near-wall region, while different mechanisms occur in the log- and wake regions. In particular, the behaviour of scale interaction away from the wall has been explained either through a preferential arrangement of the small scales -- i.e., an alignment of the small scale turbulence with internal shear layers that separate zones of large scale uniform momentum~\citep{hutchins2014large, baars2017reynolds} -- or 
as an effect of variations in the mean strain and in the shear-driven production~\citep{agostini2019departure}.
		\\Based on the insights from~\citet{hutchins2007large} and~\citet{mathis2009Large}, AM has been largely investigated for several flow configurations and Reynolds numbers, both experimentally~\citep[e.g., see ][]{mathis2009comparison, schlatter2010quantifying, guala2011interactions, ganapat2012amplitude, talluru2014amplitude, baars2015wavelet, duvvuri2015triadic, squire2016inner, baars2017reynolds, pathikonda2017inner, basley2018spatial, pathikonda2019investigation} and numerically~\citep[e.g., see ][]{chung2010Large, bernardini2011inner, agostini2014influence, hwang2016inner, agostini2016skewness, anderson2016amplitude, yao2018amplitude, dogan2019quantification, agostini2019departure}. Furthermore, findings on scale interaction have fostered the development of predictive models for near-wall turbulence that explicitly account for the footprint and amplitude modulation by large scales on small scales~\citep{marusic2010predictive, mathis2011predictive, mathis2013estimating, baars2016spectral, wu2019modelling}. It should be noted that, although large-scale spectral features do not match between internal and external flows, similar AM results have been found for channel, pipe and boundary layer flows at similar $\Rey_\tau$ values~\citep{mathis2009comparison}, suggesting a similar scale-interaction mechanism is at play in all configurations.

		Besides amplitude modulation, small scale turbulence has also been found to change its \textit{instantaneous} (i.e., local) frequency during intervals of positive or negative $u_{LS}$, namely large scales affect the smalls scales through a frequency modulation (FM) mechanism~\citep{ganapat2012amplitude, fiscaletti2015amplitude, baars2015wavelet}. However, much less investigations to quantify FM in wall-bounded turbulence have been carried out so far~\citep{ganapat2012amplitude, baars2015wavelet, baars2017reynolds, pathikonda2017inner, tang2018scale, awasthi2018numerical, pathikonda2019investigation} if compared with the vaster literature on amplitude modulation and its application into predictive models. One of the main reasons for this literature imbalance resides on the difficulty to produce robust methodologies to quantify FM in broadband signals, as well as the difficulty to effectively capture instantaneous frequencies in a signal.

		Aiming to quantify FM in the context of wall-bounded turbulence scale interaction, two methodologies have been proposed as yet. \citet{ganapat2012amplitude} proposed a peak-valley approach, following the idea that local frequency is proportional to the number of maxima and/or minima per unit length of the series. The peak-valley approach was applied to streamwise $u_{SS}$ signals from experimental measurements in a turbulent boundary layer at $\Rey_\tau=\delta U_\tau/\nu=14150$ (where $\delta$ is the boundary layer thickness). Similarly to AM, the authors found a relevant FM of the small scales in the near-wall region in which higher frequencies correspond to large (positive) $u_{LS}$ values while lower frequencies correspond to low (negative) $u_{LS}$ values. However, differently from AM, substantial FM was observed only up to $y^+\approx 100$. As an example of this FM mechanism, in \figurename~\ref{fig:egserie_SSLS}(b) a rapidly fluctuating $u_{SS}$ activity can be seen during positive $u_{LS}$ (dark blue intervals) than negative $u_{LS}$ (light blue intervals). Despite its conceptual simplicity, the main drawback of the peak-valley approach is the need of a signal discretization into sub-intervals of arbitrary spacing to quantify the number of maxima and minima within each sub-interval. The choice of the size of the signal partition into sub-intervals is non-trivial and requires a trade-off between too short or too large intervals that can affect the results. Moreover, as pointed out by~\citet{baars2015wavelet}, the short-time partitioning of the peak-valley approach makes it less applicable if temporal shifts in amplitude and frequency modulation have to be focused.

		An alternative approach to quantify FM and effectively account for time shifts was then proposed by~\citet{baars2015wavelet}, who exploited wavelet analysis to extract from the velocity time-series a new signal that is representative of the local frequency variations at the small scales. The authors performed a time-frequency analysis of the streamwise velocity, in which a time-series -- representative of the small scale instantaneous frequency -- was obtained by evaluating the first spectral moment of the wavelet power spectrum, namely an average energetic contribution at each time coming from the range of (high) frequencies pertaining the small scales~\citep{baars2015wavelet}. The first spectral moment was eventually long-wavelength pass filtered to retain only its large scale component, and correlated with $u_{LS}$ to quantify FM (similarly to the AM technique proposed by~\citet{mathis2009Large}). 
		\\The wavelet-based procedure was applied to experimental streamwise velocity time-series measured at different wall-normal locations from a turbulent boundary layer at $\Rey_\tau=14750$~\citep{baars2015wavelet}. The authors showed positive correlations up to the center of the log-region, meaning that higher and lower frequencies in $u_{SS}$ are detected under $u_{LS}>0$ and $u_{LS}<0$, respectively. Almost zero correlations were observed, instead, for higher wall-normal locations up to the boundary layer intermittent region, where negative correlation values were detected. The near-wall FM found by~\citet{baars2015wavelet} is in accordance with the outcomes from~\citet{ganapat2012amplitude}, but the $y^+$ coordinate above which FM was found to be almost absent is larger by using the wavelet-based approach ($y^+\approx 470$) than the peak-valley approach ($y^+\approx 100$), although the $\Rey_\tau$ values were rather similar. Furthermore, a phase lead of the small scale amplitude and frequency was found in the near-wall with respect to the large scale signals, and -- in accordance with previous studies~\citep{bandyopadhyay1984coupling, guala2011interactions} -- a much larger lead was detected for small scale amplitudes than for frequency. Although the FM has been accepted as a near-wall mechanism, it is still not fully clear the interaction mechanism in terms of FM between small and large scales in the log- and wake regions, in particular what is the precise wall-normal coordinate at which small scale frequency is no longer affected by large scales.

		So far, the wavelet-based technique by~\citet{baars2015wavelet} has been exploited as the main tool to quantify FM in wall-bounded turbulence. Different flow configurations have been explored in terms of FM, such as experimental smooth-wall turbulent boundary layers via hot-wire measurements~\citep{baars2017reynolds} and particle image velocimetry~\citep{pathikonda2019investigation}, experimental boundary layers in presence of wall roughness~\citep{pathikonda2017inner, tang2018scale}, as well as large eddy simulation of a turbulent channel flow with spanwise heterogeneity~\citep{awasthi2018numerical}. These works highlighted that -- despite the specific quantitative differences -- near-wall FM is present both for smooth- and rough-walls, as well as for several Reynolds numbers. However, despite its preferred employment for quantifying FM, the wavelet-based approach presents some criticalities. First, as discussed by~\citet{baars2015wavelet}, the choice of the mother wavelet can have an impact on the results since different frequency resolutions are gained from different mother wavelets. Moreover, the procedure necessitates multiple filtering operations that demand the choice of an appropriate frequency filter value. In particular, a frequency threshold is required both in the computation of the first spectral moment of the wavelet power spectrum (that involves a numerical integration), and in the long-wavelength pass filtering of the first spectral moment. Therefore, differently from the peak-valley approach by~\citet{ganapat2012amplitude} -- in which maxima and minima are counted -- the wavelet-based approach intrinsically requires several procedural steps and assumptions that need to be carefully handled.

		In this work, a novel approach to study FM in wall-bounded turbulence is put forward with a twofold aim: (i) to propose a non-parametric and robust methodology to extract local frequency changes in a signal, and (ii) to show its effectiveness for two wall-bounded turbulence configurations, also reporting novel insights that can help to further shed light on large-small scale interaction. Our methodology relies on the natural visibility graph (NVG) approach proposed by~\citet{lacasa2008}, which is used to map a signal into a network by exploiting a geometrical criterion. Thanks to its simplicity of implementation, the NVG has been widely employed in a large variety of research areas such as, among many others, economy, biomedicine, geophysics~\citep{zou2018complex}. In particular, visibility-based investigations have been carried out in fluid mechanics to study jets and fires~\citep{charakopoulos2014application, murugesan2019complex, tokami2020spatiotemporal}, wall-bounded turbulent flows~\citep{liu2010statistical, iacobello2018visibility}, passive scalar plumes~\citep{iacobello2018complex, iacobello2019experimental}, and turbulent combustors~\citep{murugesan2015combustion, murugesan2016detecting, singh2017network}.

		In spite of its simplicity, the NVG approach (defined in \S~\ref{subsec:visib_def}) has been shown to be a powerful tool in capturing important features of the mapped signal (such as the occurrence of extreme events) and a reliable indicator of the transition between different flow dynamics~\citep{iacobello2020review}. Here we show that the degree centrality -- which is one of the simplest network metrics -- is much more sensitive to the small scale spectral energy variations than the large scale counterpart (\ref{subsec:degree_features}). Accordingly, the network degree is viewed as a metric that is able to inherit the local frequency variations in a signal (\ref{subsec:degree_FM}), without any \textit{a priori} assumption (e.g., signal filtering). Therefore, the NVG approach can be directly used to study the full velocity signals rather than the small scale component.

		The proposed NVG approach is used to analyse time-series (\S~\ref{subsec:tbl}) from an experimental smooth-wall zero-pressure-gradient turbulent boundary layer~\citep[$\Rey_\tau=14750$, ][]{MARUSIC2020}, and spatial-series -- namely 1D signals along spatial transects at fixed time (\S~\ref{subsec:chan}) -- from two direct numerical simulations (DNSs) of smooth-wall incompressible turbulent channel flows~\citep[$\Rey_\tau\approx 5200$ and $\Rey_\tau= 1000$, ][]{lee2015direct, graham2016web}. In this regard, for simplicity, we refer to as FM to indicate both temporal and spatial frequency (i.e., wavenumber) modulation, where the former applies to time-series while the latter to spatial-series. A comparative FM analysis is performed by highlighting differences and similarities between outcomes from the two wall-bounded turbulence setups for the streamwise velocity (\S~\ref{subsec:res_u_FM}).  In particular, the effect of different Reynolds numbers is examined, and the application of Taylor's hypothesis to time-series is discussed by proposing a convection velocity that compensates for overprediction of modulation in the near-wall region. Moreover, FM results are examined in the view of the quasi-steady quasi-homogeneous theory, in terms of degree centrality scaling with respect to large scale velocity values (\S~\ref{subsec:res_u_QS}). The analysis is then extended to the wall-normal and spanwise velocities of the channel flow (\S~\ref{subsec:res_vw}), and time and space shifting are eventually investigated for all the three velocity components (\S~\ref{subsec:res_shifting}). Finally, we provide a discussion on some general features of the visibility approach (\S~\ref{sec:discuss}) as well as concluding remarks (\S~\ref{sec:conclusion}).

\section{Visibility-based analysis of frequency modulation}\label{sec:visib}
	
	\subsection{Definition of visibility graph}\label{subsec:visib_def}
	
\begin{figure}
	 	 \centerline{\includegraphics{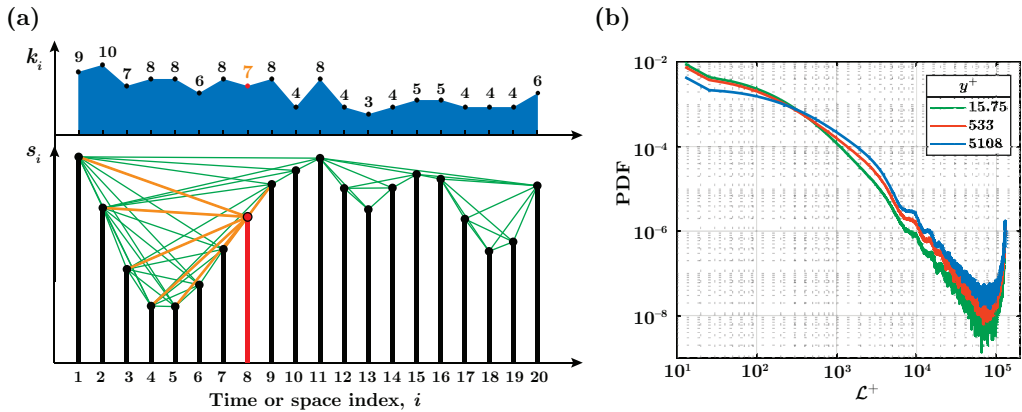}}
		  \caption{(a) The lower diagram shows an example of a signal, $s_i\equiv s(\chi_i)$, and the corresponding visibility network, where nodes are depicted as black filled circles and links as green lines. In particular, the node $i=8$ and its links are highlighted in orange. The degree values for each node, $k_i$, are also shown in the upper diagram. (b) PDF of the link length evaluated on the network built from streamwise velocity, $u(x)$, in a turbulent channel flow at $\Rey_\tau\approx 5200$. The link-length is expressed in wall-units as $\mathcal{L}^+=|i-j|\Delta x^+$, where $i$ and $j$ are the indices of two connected nodes and $\Delta x^+=12.7$ (see \S~\ref{subsec:chan}).} \label{fig:eg_visib}
		\end{figure}

		Visibility graphs represent a widely employed technique to map a discrete signal in a network. The idea behind the visibility graph approach is to assign a node of the network to each datum in the signal, and activate a link between two nodes if a geometrical criterion is satisfied. The main variant is the natural visibility graph (NVG), which is based on a convexity criterion~\citep{lacasa2008}. Geometrically, two nodes in an NVG (corresponding to two points in the signal) are linked if the straight line connecting the two points lies above any other in-between data. \figurename~\ref{fig:eg_visib}(a, lower diagram) shows an example of a short series, $s_i\equiv s(\chi_i)$, for the independent variable $\chi_i$ (i.e., a time or space coordinate), comprising $N=20$ observations, illustrated as vertical bars. Nodes and links in \figurename~\ref{fig:eg_visib}(a) are depicted as filled circles at the tip of each bar and green straight lines, respectively. A representative node is highlighted in red and its links are reported in orange. 
		\\The NVG criterion applied to a generic signal, $s(\chi)$, can be formally written as:
	
		\begin{equation}\label{eq:visib}
			s(\chi_n)<s(\chi_j)+\left(s(\chi_i)-s(\chi_j)\right)\frac{\chi_j-\chi_n}{\chi_j-\chi_i},\quad i,j=1,\dots,N,
		\end{equation}
	
		\noindent for any $\chi_n$ (i.e., time or space coordinate) such that $\chi_i<\chi_n<\chi_j$~\citep{lacasa2008}. The corresponding visibility network is represented through the adjacency (binary) matrix $\boldsymbol{A}$, whose entries are $A_{i,j}=1$ if the inequality~(\ref{eq:visib}) is satisfied for the node pair $\left(i,j\right)$ with $i\neq j$, and $A_{i,j}=0$ otherwise. For example, in \figurename~\ref{fig:eg_visib}(a), the node $i=8$ is connected (i.e., $A_{8,j}=1$) to nodes $j=\lbrace 1, 2, 3, 4, 5, 7, 9\rbrace$, as highlighted by the orange links. By definition, visibility networks are connected (i.e., each node $i$ is linked to at least one other node $j$, e.g., $j=i+1$ or $j=i-1$) and undirected~\citep{newman2018networks}, namely the adjacency matrix is symmetric ($A_{i,j}=A_{j,i}$). 
			
		Differently from other techniques developed to transform a signal into a network~\citep{zou2018complex, iacobello2020review}, the visibility algorithm does not require any \textit{a priori} parameter. Given a signal, a unique visibility network is obtained in a straightforward way by applying the convexity criterion in~(\ref{eq:visib}) for each pair of data. Another feature of NVGs is the invariance under affine transformations of the mapped signal, namely translation and rescaling (i.e., multiplication by a positive constant) of both horizontal and vertical axes~\citep{lacasa2008}. This implies that two signals with the same temporal (or spatial) structure but with different mean values (i.e., vertical translation of the series) and standard deviations (i.e., vertical rescaling of the series) are mapped in the same visibility graph. 
		
		In the present work, we exploited the NVG approach to study turbulent velocity signals from wall-bounded turbulence, both as time-series (from the boundary layer, \S~\ref{subsec:tbl}) and spatial-series (from the channel flow, \S~\ref{subsec:chan}). We note that this is the first time the NVG is employed for studying wall-bounded turbulence by focusing on spatial-series rather than time-series. An optimized code for computing the NVG (either for spatial- and time-series) was provided by~\citet{IacobelloNVGmatlab}, where the possibility to account for spatial-series periodicity is also implemented.

		One remarkable feature of NVGs from signals referring to physical phenomena with a wide range of different scales (such as in turbulence) is the infrequent appearance of long-range links. In fact, the presence of fluctuations of different amplitude in the signal prevents the possibility that a node is visible by other distant nodes~\citep{zhuang2014time}. To grasp the concept, the probability density function (PDF) of the link length in $u(x)$ signals from a turbulent channel flow ($\Rey_\tau\approx 5200$, see \S~\ref{subsec:chan}) is shown in \figurename~\ref{fig:eg_visib}(b), evidencing that long-range links are very unlikely to occur (the increasing PDF for large $\mathcal{L}^+$ values is due to signal periodicity in the $x$-direction).

	The capability of visibility graphs to capture the temporal (or spatial) structure of a signal by means of a convexity-based geometrical framework, hence, turns out to be a key feature to study the occurrence in time (or space) of specific events~\citep{iacobello2018complex, iacobello2019experimental}. In this work, we take advantage from the features of visibility networks to detect frequency modulation of large scales on small scales.

	\subsection{Node degree in relation to small scale signal features}\label{subsec:degree_features}
				
		 The degree centrality (or, simply, degree) of a node, $i$, is defined as the number of \textit{neighbours} of $i$, that is the number of nodes linked to $i$, 
		
		\begin{equation}\label{eq:degree}
			k_i\equiv\sum_{j=1}^N{A_{i,j}},
		\end{equation}
		
		\noindent where $N$ is the total number of nodes, corresponding to the number of sampled values of the signal~\citep{newman2018networks}. The top panel in \figurename~\ref{fig:eg_visib}(a) shows the sequence of degree values for the example of signal, $s_i$, shown in the bottom of \figurename~\ref{fig:eg_visib}(a); for instance, the degree of node $i=8$ (highlighted in red) is $k_8=7$ since it is connected to seven other points (links are highlighted in orange). By averaging over all nodes, a representative degree value for the network (i.e., for the whole signal) is obtained as $K=\sum_{i}{k_i}/N$. It should be noted that the degree, $k_i$, provides a measure of the extent to which a single node $i$ belongs to a convex interval in the signal, but it is not directly able to quantify whether the properties of node $i$ (e.g., its importance in the network) are similar or not to the properties of other nodes. Instead, this issue can be tackled through assortativity measures, which can be used to assess similarities among nodes (e.g., in terms of their importance in the network through degree-degree correlation)~\citep{newman2018networks}.

		\begin{figure}
	 		\centerline{\includegraphics{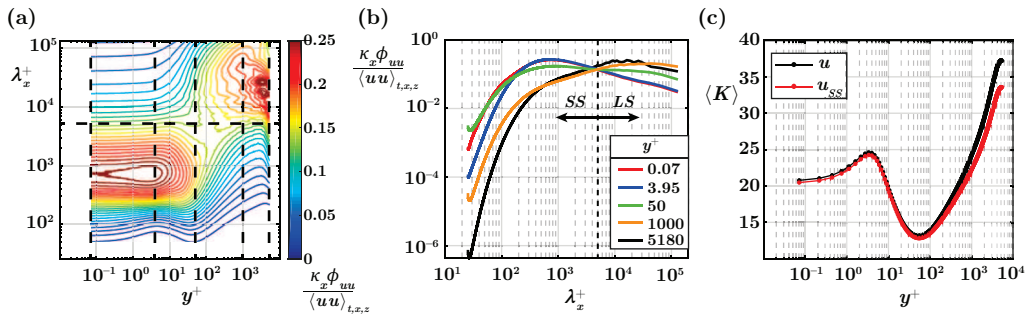}}
		  	\caption{(a) Pre-multiplied energy spectral density, $\phi_{uu}$, from a turbulent channel flow at $\Rey_\tau\approx5200$ (see \S~\ref{subsec:chan}), normalized by the streamwise velocity variance, $\langle uu\rangle_{t,x,z}$. The horizontal dashed line indicates the value of spectral filter, while vertical dashed lines highlight five representative $y^+$ coordinates. (b) Pre-multiplied energy spectral density for the five selected $y^+$ locations in (a). (c) Wall-normal behaviour of the average degree centrality, $\langle K\rangle$, for NVG built from the full streamwise velocity, $u(x_i)$ (black curve), and from the small scale streamwise velocity, $u_{SS}(x_i)$ (red curve), obtained through a spectral decomposition. Angular brackets in $\langle K\rangle$ indicate averaging over the (homogeneous) spanwise direction, $z$.} \label{fig:spectra_deg}
		\end{figure}

		Recalling that long-range links unlikely appear in visibility graphs (\figurename~\ref{fig:eg_visib}(b)), the main contribution to degree values is due to short-range links, making the degree a metric that is typically sensitive to the local structure of the signal. Rapidly fluctuating signals are then expected to show lower degree values, $k_i$, and in turn a lower average degree $K$~\citep{zhuang2014time}. Since rapid variations in the local structure of turbulent signals are mainly governed by high frequencies (i.e., low wavelengths), a relation should exist between the average degree, $K$, and the high-frequency spectral energy (that produces the local variations in the turbulent signals).
	
		With the aim to explore this relation, we report in \figurename~\ref{fig:spectra_deg}(a) the energy spectral density of the streamwise velocity, $\phi_{uu}$, pre-multiplied for the wavenumber, $\kappa_x=2\pi/\lambda_x$, from a turbulent channel flow at $\Rey_\tau\approx 5200$ (see \S~\ref{subsec:chan}). Notice that $\phi_{uu}$ is normalized by the variance of the streamwise velocity fluctuations, $\langle uu\rangle_{t,x,z}$ (here angular brackets indicate the average over time, $t$, and homogeneous directions, $x,z$). In \figurename~\ref{fig:spectra_deg}(a) it is easily distinguishable the spectral peak separation between small and large scales, as well as the spectral filter adopted in this work, marked as a horizontal dashed line. Five curves of the spectrum at five representative $y^+$ coordinates (highlighted as dashed vertical lines in \figurename~\ref{fig:spectra_deg}(a)) are also shown in \figurename~\ref{fig:spectra_deg}(b). 
		\\The rationale behind the normalization of the spectrum through the variance is twofold: on the one side, the streamwise energy density at each $y^+$ is accentuated, thus emphasizing the occurrence of the two spectral peaks and, on the other side, this normalization permits a congruent comparison with the degree behaviour computed on visibility networks (which are insensitive to different variance levels, i.e., on signal rescaling). Moreover, due to the variance normalization, the area under each curve in \figurename~\ref{fig:spectra_deg}(b) is equal to unity, so that the integral of curves in \figurename~\ref{fig:spectra_deg}(b) in a given range of $\lambda_x$ represents the fraction of total energy pertaining that scale range. In this way, \figurename~\ref{fig:spectra_deg}(b) elucidates the redistribution of the spectral energy density over scales, $\lambda_x^+$, at different $y^+$ coordinates (a log-log plot in \figurename~\ref{fig:spectra_deg}(b) is shown with the aim to highlight the behaviour at small $\lambda_x^+$ values).

		Focusing on the small scales (say, $\lambda_x^+<\Rey_\tau\approx5200$) in \figurename~\ref{fig:spectra_deg}(b), we observe that by moving from very close to the wall ($y^+\approx 0.07$) up to $y^+\approx 4$ there is a small decrease in the (normalized) energy content, then an increase of the (normalized) spectral energy occurs from $y^+\approx 4$ up to the beginning of the log-layer ($y^+\approx 50$), and lastly a persistent decrease happens up to the channel centreline ($y^+\approx 5200$). A reduction or a growth of the (normalized) spectral energy at small scales indicates that the signal tends to be locally smoother (i.e., slowly-varying, without rapid low-intensity fluctuations) or more irregular (i.e., rapidly-varying), respectively. Recalling that the mean degree, $K$, is sensitive to the local structure of the signal, an increase of the degree values is then expected for locally smoother signals (i.e., low spectral energy at local scales), and \textit{vice versa}. \figurename~\ref{fig:spectra_deg}(c) shows the wall-normal behaviour of the mean degree, $K$, of networks built from the full streamwise velocity, $u(x_i)$ (black line), in the channel flow setup. As expected, the $y^+$-trend of $K$ for the full signal closely follows the behaviour of the small scale spectral energy density as described above, where the degree growth is faithfully related to the small scale spectral-energy decrease, and \textit{vice versa}. In particular, we point out that the value of $K$ at each $y^+$ is associated to an integral effect of all wavelengths in the signal, so that $K(y^+)$ is due to a cumulative effect of different spectral-energy levels.
		\\ \figurename~\ref{fig:spectra_deg}(c) also shows the $y^+$-behaviour of the mean degree of networks built from $u_{SS}(x_i)$ (red line), namely, in which the large scale component is removed. The values and the trends of $K$ from the full and the small scale velocity signals are very close, and a slight difference appears only very far from the wall. Note that a similar behaviour of $K$ as that shown in \figurename~\ref{fig:spectra_deg}(c) for the channel case is also found for the turbulent boundary layer case. It should be noted that, since very long-range connections are unlikely to appear (\figurename~\ref{fig:eg_visib}(b, they only barely contribute to the average degree, $K$, which instead is mainly related to shorter links.
		
		The very good agreement between $K(y^+)$ for the full $u$ signal and the $y^+$-variations in the small scale spectral energy (corroborated by the similarity of $K(y^+)$ for $u$ and $u_{SS}$) indicates that the network degree is able to capture the features of the small scale turbulence directly from the full signal, i.e. without the arbitrary requirements of filtering operations. These features will be exploited in the next Section~\ref{subsec:degree_FM} to provide a metric which is able to quantify frequency modulation. We notice that, to the best of our knowledge, this is the first time that insights from the visibility graph approach are directly related to spectral properties of a signal.

	\subsection{FM detection via degree centrality}\label{subsec:degree_FM}

\begin{figure}
		  \centerline{\includegraphics{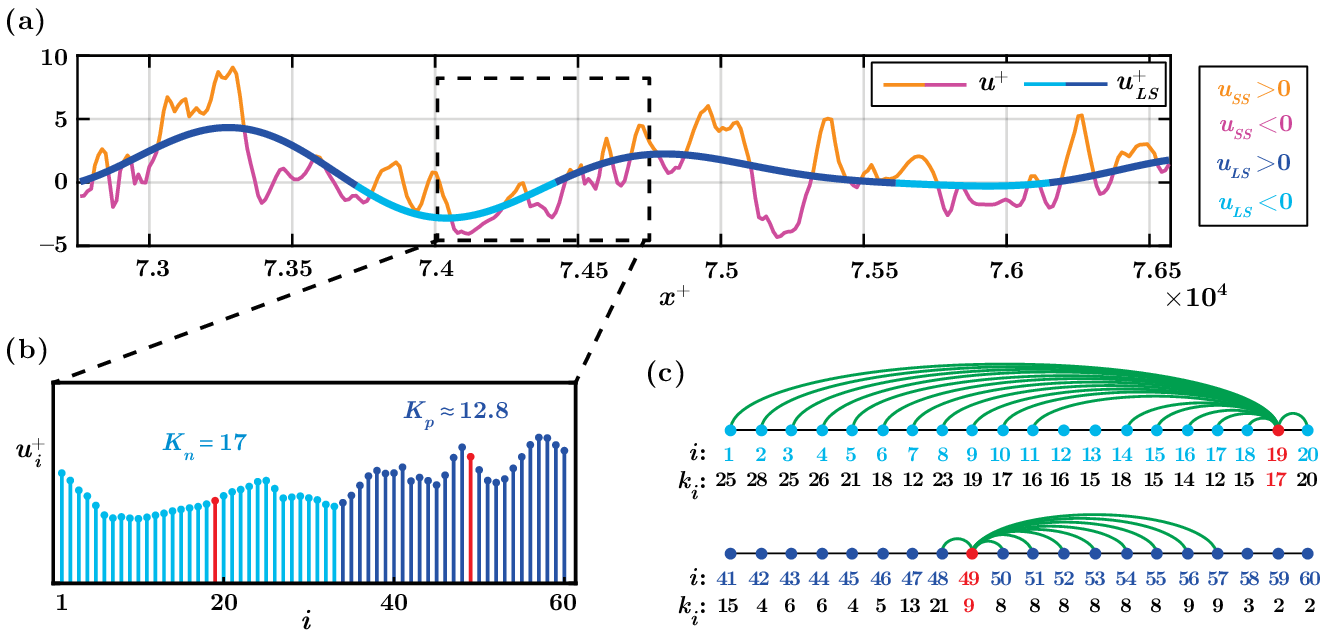}}
		  \caption{(a) An interval of the streamwise velocity, $u$, and its large scale component, $u_{LS}$, extracted along the streamwise direction, $x$, of the turbulent channel flow at $y^+\approx 10$. The full $u$ signal is depicted as orange-magenta lines indicating intervals of positive and negative small scale velocity fluctuations, $u_{SS}=u-u_{LS}$, respectively, while the $u_{LS}$ signal is depicted as light-dark blue lines, highlighting intervals of $u_{LS}<0$ and $u_{LS}>0$, respectively. Both the series are normalized in wall-units. (b) A piece consisting of $60$ data of the velocity $u$ from panel (a), depicted as vertical bars whose color reflects the sign of $u_{LS}$. Two representative data (corresponding to nodes $i=19$ and $i=49$ of the NVG network) are highlighted in red. (c) Network representation of the NVG built from signal in panel (b). Two subsets of nodes and links from the two representative nodes, $i=\lbrace 19, 49\rbrace$, are shown as coloured dots and green arcs, respectively. The sequence of degree values, $k_i$, for each node, $i$, is also reported.}\label{fig:deg_ist_per}		  
		\end{figure}

		The aim of this section is to provide a degree-based metric able to quantify FM from full velocity signals. With this aim, in \figurename~\ref{fig:deg_ist_per}(b) we show a short representative interval of the streamwise velocity series reported in \figurename~\ref{fig:deg_ist_per}(a), which is extracted from the turbulent channel flow at $y^+\approx 10$. The corresponding NVG is then built from the short signal in \figurename~\ref{fig:deg_ist_per}(b), and the links activated by two representative nodes, $i=\lbrace 19,49\rbrace$ (highlighted as red dots in \figurename~\ref{fig:deg_ist_per}(b)), are shown as green arcs in \figurename~\ref{fig:deg_ist_per}(c). Node $i=19$ clearly displays more connections than node $i=49$ (i.e., $k_{19}>k_{49}$), since node $i=19$ is in a larger convex interval than $i=49$; in other words, the signal around $i=49$ varies more rapidly than around $i=19$. Therefore, although the degree, $k_i$, represents a pointwise value because $k_i$ refers to a single coordinate $i$, the information enclosed in $k_i$ originates from the surroundings of $i$. The degree $k_i$ can then be interpreted as a measure of the \textit{instantaneous period} (or \textit{instantaneous wavelength}) at the temporal (or spatial) coordinate $t_i$ (or $x_i$), in analogy with the concept of instantaneous frequency used in signal analysis~\citep{huang1998empirical, boashash2015time}. Larger $k_i$ values correspond to larger instantaneous periods (or wavelengths), and in turn to smaller instantaneous frequencies.
		
		On the basis of this argument and the insights illustrated in the previous Section~\ref{subsec:degree_features}, we introduce the ratio, $K_{np}$, to quantify frequency modulation, defined as
		
		\begin{equation}\label{eq:dndp}
			K_{np}\equiv\frac{K_n}{K_p}, \qquad K_n=\frac{1}{N_{neg}}\sum_{j=1}^N{(k_j|u_{LS}<0)}, \qquad K_p=\frac{1}{N_{pos}}\sum_{j=1}^N{(k_j|u_{LS}>0)},
		\end{equation}
		
		\noindent where $K_n$ and $K_p$ are the average degree values computed on the NVG of the full velocity signal, conditioned to intervals of $u_{LS}<0$ and $u_{LS}>0$, respectively, while $N_{neg}$ and $N_{pos}$ are the number of occurrences in which $u_{LS}<0$ and $u_{LS}>0$, respectively.
		\\Values of $K_{np}$ greater than $1$ indicate that the degree is (on average) larger during $u_{LS}<0$ intervals than during during $u_{LS}>0$, and \textit{vice versa} for $K_{np}$ smaller than $1$. In the example of \figurename~\ref{fig:deg_ist_per}(b-c), the degree values, $k_i$, of the two representative nodes in red ($k_{19}=17$ and $k_{49}=9$) exemplify the behaviour of the two signal intervals during $u_{LS}<0$ and $u_{LS}>0$, resulting in $K_n=17$, $K_p\approx 12.8$ and $K_{np}>1$. Hence, $K_{np}$ discriminates between (i) positive frequency modulation for $K_{np}>1$ (i.e., an increase of the local frequency in the velocity signal gained for $u_{LS}>0$ and a decrease for $u_{LS}<0$), (ii) negative frequency modulation for $K_{np}<1$, and (iii) an absence of modulation for $K_{np}\approx 1$. We emphasize that the arguments leading to the ratio~(\ref{eq:dndp}) do not involve any \textit{a priori} parameter, but the unique availability of the full velocity signal to compute the degree value of each node. A filtering operation is only required to condition the degree values to the sign of the large scale velocity.

		To test the NVG-based approach, we built synthetic signals that mimic the near-wall modulation mechanism in wall-bounded turbulence for three modulation cases: amplitude modulation (AM), frequency modulation (FM), and both amplitude and frequency modulation. Appendix~\ref{app_synt} contains details on the synthetic signal construction and reports the $K_{np}$ values for each configuration (see \figurename~\ref{fig:eg_synt_FM}), showing that $K_{np}$ is able to highlight the presence of FM and -- in presence of both AM and FM mechanisms -- tends to be more sensitive to FM while only weakly to AM.

		In summary, the $K_{np}$ ratio combines the capability of visibility networks (i) to capture the information on the local temporal structure of a series (\S~\ref{subsec:visib_def}), and (ii) to inherit the small scale energetic features from the full signal (\S~\ref{subsec:degree_features}). These characteristics make the visibility approach a powerful and easy-to-use alternative to previously proposed methodologies for time-frequency characterization of turbulence signals. In the following, the NVG-approach is carried out for wall-turbulent signals, showing its robustness (with respect to different cut-off filtering size) and effectiveness in capturing the large-to-small scale FM mechanism.

\section{Description of the turbulent flow datasets}\label{sec:datasets}

	Two main datasets of high Reynolds number wall-bounded turbulent flows are exploited in this work to study frequency modulation by means of visibility network-based tools: (i) spatial-series from a numerically-simulated turbulent channel flow at $\Rey_\tau\approx 5200$~\citep{lee2015direct}, and (ii) time-series from experimental measurements in a turbulent boundary layer at $\Rey_\tau=14750$~\citep{MARUSIC2020}. Although outer flow structures start to occur and play a role in scale interaction at lower Reynolds numbers~\citep{agostini2014influence, hu2018energy, wu2019modelling}, high Reynolds number flows are required to enhance the inter-scale separation and amplify the scale interaction mechanism. Moreover, a third DNS dataset of turbulent channel flow at $\Rey_\tau=1000$ is also employed for comparison purposes, thus showing effects of inertial on FM results.
	\\To the best of our knowledge, this is the first time a state-of-the-art DNS at $\Rey_\tau\approx 5200$ is employed to specifically investigate large-to-small scale FM. In fact, while high Reynolds number boundary layer flows are typically obtained in experimental facilities (as witnessed by most of previous works on AM and FM), high-$\Rey_\tau$ experiments of channel flows are difficult to realise due to strong side-wall boundary effects~\citep{lee2015direct}. The DNS employed in this work is at a large enough Reynolds number~\citep[i.e., $\Rey_\tau >4000$, as reported by ][]{hutchins2007large} to guarantee a sufficient large-small scale spectral separation (e.g., see energy peaks separation in \figurename~\ref{fig:spectra_deg}(a)), and allows us to perform a FM analysis on all the three velocity components that, so far, has only been performed for AM~\citep[e.g., see ][]{talluru2014amplitude, agostini2016predicting}.

		The scale decomposition of the streamwise velocity fluctuation signals was performed as $u(x)=u_{LS}(x)+u_{SS}(x)$ (e.g., \figurename~\ref{fig:deg_ist_per}(a)) and $u(t)=u_{LS}(t)+u_{SS}(t)$ (e.g., \figurename~\ref{fig:egserie_SSLS}(b)) for the spatial- and time-series taken from the turbulent channel and boundary layer flows, respectively. A common approach to obtain $u_{SS}$ and $u_{LS}$ is to employ a spectral filter to retain the high and low wavelength or frequency, respectively, as performed in several previous works~\citep{mathis2009Large, ganapat2012amplitude, baars2015wavelet, baars2017reynolds, pathikonda2017inner, pathikonda2019investigation}. Alternatively,~\citet{agostini2014influence} proposed to employ the \textit{empirical mode decomposition}~\citep{huang1998empirical} to separate large and small scale contributions. In this work, both the spectral and empirical mode decompositions were tested to separate the large and small scale contributions. However, for the sake of simplicity and in line with most of the current literature, results are only shown for a spectral decomposition, as both the procedures produce equivalent results.

	\subsection{DNS of turbulent channel flows}\label{subsec:chan}

		Velocity fields were extracted from two direct numerical simulations of incompressible turbulent channel flows. The first DNS was run at frictional Reynolds number $\Rey_\tau\equiv h U_\tau/\nu=5186$, where $h=1$ is the half-channel height, $U_\tau=4.14872\times 10^{-2} U_b$ and $\nu=8\times 10^{-6} U_b h$, with the bulk velocity $U_b=1$. The size of the spatial domain is $\left(8\pi h \times 2h \times 3\pi h\right)$ with $\left(10240 \times 1536 \times 7680\right)$ grid points along the streamwise, wall-normal and spanwise directions, respectively. The flow fields were recorded only after statistical stationarity of the flow was reached, and $11$ temporal frames of velocity and pressure spatial fields were stored in the dataset. The time interval between two consecutive frames is about $0.7$ flow-through time, corresponding to about $3785\nu/U_\tau^2$ in wall-units. 
		\\The simulation was performed at a sufficiently high Reynolds number and with a sufficiently large spatial domain to exhibit characteristics of high-Reynolds-number turbulence, e.g., the presence of large scale motions and a rather large wall-normal range for statistics scaling~\citep{lee2015direct}. The dataset is available online (doi:\href{https://doi.org/10.7281/T1PV6HJV}{10.7281/T1PV6HJV}) from the Johns Hopkins Turbulence Database~\citep{li2008public}. For further simulation details and statistics, see~\citet{lee2015direct}.
		
		The second DNS was run at $\Rey_\tau\equiv h U_\tau/\nu=1000$, with $h=1$, $U_\tau=4.9968\times 10^{-2} U_b$, $\nu=5\times 10^{-5} U_b h$ and $U_b=1$. The size of the spatial domain is $(8\pi h \times 2h \times 3\pi h)$ with $(2048 \times 512 \times 1536)$ grid points along the streamwise, wall-normal and spanwise directions, respectively. Data were stored for approximately one flow-through time, $ [0,26]h/U_b$, with a storage temporal step of $0.0065$. Also this dataset is available online (doi:\href{https://doi.org/10.7281/T10K26QW}{10.7281/T10K26QW}) from the Johns Hopkins Turbulence Database~\citep{li2008public}. For further simulation details, see~\citet{graham2016web}.

		In this work, 1D spatial-series (i.e., extracted at a fixed time) of the three velocity components, $u,v,w$, along the streamwise direction, $x$, are exploited to build visibility networks. Network-based results are averaged in time (i.e., on $11$ temporal frames for the $\Rey_\tau\approx 5200$ setup, and on $400$ uniformly spaced temporal frames for the $\Rey_\tau= 1000$ setup) and in the spanwise direction; in the latter case, averages are performed for a set of uniformly-spaced spanwise locations separated from each other by $64$ and $128$ grid points for the $\Rey_\tau\approx 5200$ and $\Rey_\tau=1000$ configurations, respectively.
		\\The cut-off spectral filter to separate large and small scale streamwise velocity for the $\Rey_\tau\approx 5200$ setup is set equal to $\lambda_{x,c}=h$ (i.e., $\lambda_{x,c}^+=5186$), in analogy with previous works in which $\lambda_{x,c}$ is set equal to the boundary layer thickness~\citep{hutchins2007large, mathis2009Large, mathis2009comparison, marusic2010predictive, dogan2019quantification, wu2019modelling}. For the $\Rey_\tau=1000$ setup, the cut-off filter is $\lambda_{x,c}^+=5000$, thus being comparable with $\lambda_{x,c}^+$ of the higher-$\Rey_\tau$ channel flow setup.

	\subsection{Experimental turbulent boundary layer at $\Rey_\tau\approx 14750$}\label{subsec:tbl}
	
		Experimental measurements were performed in the wind-tunnel facility of the University of Melbourne, which employs a $27\,\mathrm{m}$ test section, under a free-stream velocity $U_\infty=19.95\,\mathrm{m/s}$~\citep{baars2015wavelet}. Under these conditions, a zero-pressure-gradient boundary layer develops at a frictional Reynolds number $\Rey_\tau\equiv \delta U_\tau/\nu=14750$, where $\delta=0.361\,\mathrm{m}$ is the boundary layer thickness at the measuring location (i.e., $21.65\,\mathrm{m}$ from the inlet of the test section), while $U_\tau=0.626\,\mathrm{m/s}$ and $\nu=1.532\times 10^{-5}\,\mathrm{m^2/s}$ at the same streamwise location. The dataset is the same employed by \citet{baars2015wavelet, baars2017reynolds}, and is available online at the Fluid Mechanics Research webpage of the University of Melbourne~\citep{MARUSIC2020}.
		\\Time-series of the streamwise velocity were simultaneously recorded by means of two constant-temperature hot wire probes, one at a fixed wall-normal location at $y^+=4.33$, and the other vertically moved throughout the boundary layer in the range $y^+\in[10.5,2.14\times10^4]$ (or $y/\delta\in[7.087\times10^{-4},1.45]$) for $40$ vertical locations. At each wall-normal measuring location, three sets of data are recorded at a sampling frequency of $20\,\mathrm{kHz}$, each one for $120\,\mathrm{s}$ corresponding to a large scale time $6.6\times10^3\delta/U_\infty$, thus ensuring the convergence of spectral statistics at the longest energetic wavelengths~\citep{baars2015wavelet}. The resulting time-step in wall units is $\Delta t^+=1.28$. Further details on the measurement procedure and instrumentation can be found in~\citet{baars2015wavelet}.
		
		In order to separate large and small scale components, we employ a cut-off spectral filter $\lambda_{x,c}^+=7000$ following~\citet{hutchins2007large},~\citet{hutchins2014large} and~\citet{baars2015wavelet, baars2017reynolds}, who showed this is a proper filter value for turbulent boundary layers at high Reynolds numbers. Differently from the channel flow setup in which spatial-series are considered, here the spectral filter is converted in terms of frequency by invoking the Taylor's hypothesis as $f_c(y^+)=U_c(y^+)/\lambda_{x,c}$, where $U_c(y^+)$ is a local convection velocity at the wall-normal coordinate $y^+$. The effects of different convection velocities on FM will be elucidated in the next \S~\ref{sec:results}, where a comparison with spatial data from DNSs is carried out.

\section{Results}\label{sec:results}

	The results of the application of the degree centrality as a metric to quantify frequency modulation are reported in this section, for velocity signals extracted from the turbulent channel flows and the turbulent boundary layer described above. A one-point modulation analysis is carried out: the large scale component, $u_{LS}$, used to condition the degree on the $u_{LS}$ sign (see equation~\ref{eq:dndp}) is extracted at the same $y^+$ in which the signal is mapped into a visibility network. Due to the footprint of the large and very-large scale motions towards the wall, $u_{LS}$ evaluated at each $y^+$ represents a good estimate of the large scale velocity component in the outer region, thus resulting in a more applicable procedure than two-points analysis~\citep{mathis2009Large}. In fact, two-point synchronized measurements are not easy to perform experimentally~\citep{mathis2009Large}. Previous works have shown that similar results are obtained by adopting a one- or two-point procedure for characterizing scales interaction~\citep[see, among others, ][]{hutchins2007evidence, mathis2009Large, ganapat2012amplitude}, thus one-point modulation is here preferred for simplicity.

	First, the streamwise velocity component, $u$, is considered, both in an overall perspective (\S~\ref{subsec:res_u_FM}) and with near-wall focus (\S~\ref{subsec:res_u_QS}). Most of the current literature on scale interaction in wall-bounded turbulence is indeed focused on the $u$ component, being the component in which large and small scales can be clearly separated. We then extend the analysis to the other velocity components, $v$ and $w$ (\S~\ref{subsec:res_vw}). Finally, a space-shifted FM analysis is carried out for all velocity components and for both the turbulence configurations (\S~\ref{subsec:res_shifting}).

	\subsection{FM in the streamwise velocity component}\label{subsec:res_u_FM}
	
\begin{figure}
		  \centerline{\includegraphics{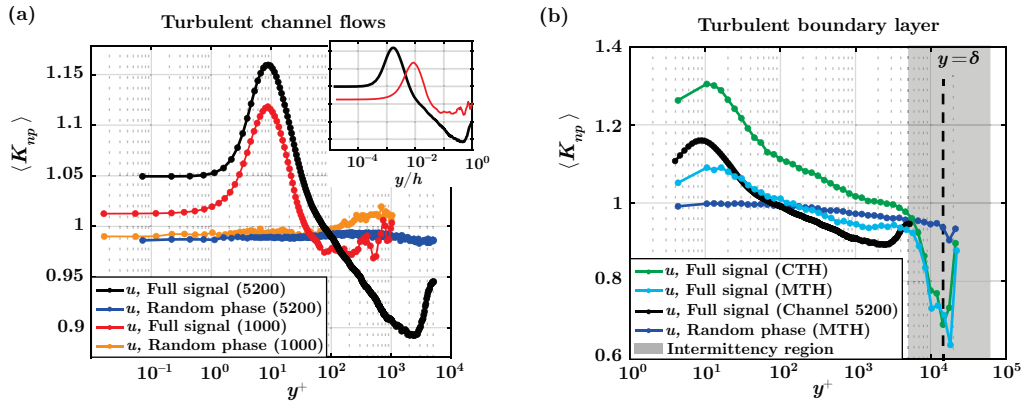}}
		  \caption{Large scale conditional average degree ratio, $K_{np}$, as a function of the wall-normal coordinate, $y^+$, for streamwise velocity, $u$, extracted from (a) the channel flow DNS at different $\Rey_\tau$ and (b) the boundary layer experiment. The inset in (a) shows the $K_{np}$ behaviour for the two channel flows as a function of $y/h$. In panel (b), $K_{np}(y^+)$ is shown for spatial-series obtained through the classical (CTH) and modified (MTH) Taylor's hypothesis as green and cyan plots, respectively; for comparison, the behaviour for the channel flow at $\Rey_\tau\approx 5200$ is also reported in black. Moreover, the boundary layer intermittency region is highlighted in (b) as a shaded grey region. Angular brackets indicate averaging over time and spanwise direction in (a) and over three different realizations in (b). The results for synthetic velocity signals with shuffled phases are also shown.} \label{fig:res_knp_U}
		\end{figure}

		The values of the ratio $K_{np}$ as a function of the wall-normal coordinate, $y^+$, are shown in \figurename~\ref{fig:res_knp_U} for visibility networks built from the streamwise velocity, $u$. We recall that $K_{np}>1$ indicates a higher frequency under intervals of positive large scale velocity than under negative ones, and \textit{vice versa} for $K_{np}<1$. 
\\\figurename~\ref{fig:res_knp_U}(a) shows $K_{np}$ for the spatial-series of the two channel flow DNS at $\Rey_\tau\approx 5200$ (black) and $\Rey_\tau=1000$ (red), while \figurename~\ref{fig:res_knp_U}(b) illustrates the $K_{np}$ behaviour for time-series of the boundary layer (green). Values of $K_{np}>1$ are detected close to the wall for all configurations, while -- moving away from the wall -- $K_{np}$ becomes smaller than $1$, indicating a reverse scale interaction mechanism, i.e., higher frequency are detected during $u_{LS}<0$ than under $u_{LS}>0$. The overall behaviours shown in \figurename~\ref{fig:res_knp_U} are in accordance with previous works on scale interaction in wall-bounded turbulence, which have indicated that a higher (amplitude and) frequency of the small scales is found under positive $u_{LS}$ intervals in regions close to the wall, while a reversal mechanism occurs far from the wall~\citep{ganapat2012amplitude, baars2015wavelet, baars2017reynolds, pathikonda2017inner, tang2018scale, awasthi2018numerical, pathikonda2019investigation}. However, the behaviour of $K_{np}(y^+)$ for the two setups highlights also peculiar features of large-to-small scale FM that deserve further investigations.

	First, we compare the results for the two channel flows at different Reynolds numbers (red and black lines in \figurename~\ref{fig:res_knp_U}(a)). A similar trend of $K_{np}(y^+)$ is found for both channels, but the intensity of the FM (close to the wall) is larger for $\Rey_\tau\approx 5200$ than for $\Rey_\tau=1000$, thus clearly showing the effect of higher Reynolds numbers is to increase FM mechanism in the near-wall region, as a consequence of increasing magnitude of the large-scale fluctuations with increasing $\Rey_\tau$. Similarly, away from the wall the reversal mechanism of scale interaction is strengthened for the higher Reynolds number DNS. Furthermore, in \figurename~\ref{fig:res_knp_U}(a) -- and also in \figurename~\ref{fig:res_knp_U}(b) for the boundary layer --  we observe a peak of $K_{np}$ at $y^+\approx 10$: this peak turns out to be related to strong sweep-like events, a phenomenon referred to as \textit{splatting} in which large scales transport high-intensity small scales towards the wall below the buffer layer~\citep{agostini2014influence, agostini2016skewness}. Thus, being the highest small scale intensity detected in the buffer layer~\citep{agostini2016skewness}, the strongest FM is revealed by a peak in $K_{np}$, which then represents a sensitive metric to local changes in the flow dynamics. Eventually, it is remarkable to observe in \figurename~\ref{fig:res_knp_U}(a) the near-wall agreement of $K_{np}$ plotted against $y^+$ between the two channel flows at different Reynolds numbers, as the near-wall dynamics is related to near-wall cycle whose characteristic scales are fixed in wall-units (see also a discussion on characteristic near-wall spatial and temporal scales in \S~\ref{subsec:res_shifting}).

	The behaviour of $K_{np}$ obtained from time-series of the turbulent boundary layer is shown in \figurename~\ref{fig:res_knp_U}(b). When time-series are considered, a convection velocity, $U_c$, has to be defined to apply Taylor's hypothesis in filtering large and small scales. Typically, $U_c$ is set equal to the local mean velocity, $U$, and in the following this assumption will be referred to as classical Taylor's hypothesis (CTH). $K_{np}$ as a function of $y^+$ obtained through the CTH is displayed in green in \figurename~\ref{fig:res_knp_U}(b): although a similar behaviour with respect to the channel flows is observed (e.g., the black line in \figurename~\ref{fig:res_knp_U}(b)), there is a significant vertical shift when time-series are employed. It is worth noting that, since the local mean velocity, $U$, does not depend on time, the temporal structure of the time-series is preserved when classical Taylor's hypothesis is applied. This implies that the application of any technique (including NVG) to study scale-interaction from time-series is the same as from the corresponding spatial-series (i.e., obtained through the classical Taylor's hypothesis), being $\Delta x\propto \Delta t$.

	The overestimation of modulation parameters when time-series and CTH are used has been previously observed for amplitude modulation in jet~\citep{fiscaletti2015amplitude}, mixing layer~\citep{fiscaletti2016scale} and turbulent boundary layer flows~\citep{yang2018implication}. Specifically,~\citet{yang2018implication} reported a distortion of the spatial-series when the classical Taylor's hypothesis is used, and suggested to employ a convection velocity, $U_c$, defined as $U_c(t)=U+\alpha u(t)$, where $u(t)=u_{LS}(t)+u_{SS}(t)$ is the fluctuating component of the streamwise velocity, and $\alpha=O(1)$ is a proportionality constant. The correction proposed by \citet{yang2018implication} is based on the rationale that the sampling time step has to be scaled using the local viscous scales, so that small scale activity is enhanced (or reduced) where the wall shear stress is high (or low) due to an increase (decrease) in the local friction velocity~\citep{yang2018implication}. However, variations in the (fluctuating) wall shear stress are mainly induced by variations into large scale fluctuations, $u_{LS}(t)$, rather than $u(t)$, as observed by~\citet{yang2018implication} and reported in previous literature~\citep[e.g., ][]{zhang2016quasisteady, baars2017reynolds} (see a more detailed discussion about the relation between near-wall small scales and wall shear stress in \S~\ref{subsec:res_u_QS}). Therefore, in this work we exploit the time-varying formulation by~\citet{yang2018implication} but only accounting for the large scale component of $u(t)$, namely

	\begin{equation}\label{eq:MTH}
		U_c(t)=U+\alpha u_{LS}(t).
	\end{equation}

	\noindent In what follows, we will refer to the application of (\ref{eq:MTH}) as the convection velocity as the modified Taylor's hypothesis (MTH), where we selected $\alpha=0.8$ (which has proved to be a suitable value). It should be noted that a correction based on $u_{LS}$ arguments was also discussed by~\citet{fiscaletti2015amplitude} for jet and boundary layer flows to compensate for amplitude modulation overestimation. Moreover, it should be noted that, differently form the classical Taylor's hypothesis, the structure of $u(x)$ is different than the structure of $u(t)$ for the MTH case, since $U_c(t)$ is not a constant thus leading to a non-uniform spacing of the spatial series.
	
	The $K_{np}$ behaviour for the MTH is shown in cyan in \figurename~\ref{fig:res_knp_U}(b): the overestimation of $K_{np}$ is compensated and its values are much more comparable to spatial-series obtained from DNS of the channel flow at $\Rey_\tau\approx 5200$ (black line in \figurename~\ref{fig:res_knp_U}(b)). This analysis confirms the applicability of the time-dependent correction of~\citet{yang2018implication} for amplitude modulation when time-series are employed, and extends such correction to the study of frequency modulation through (\ref{eq:MTH}). In particular, we stress that $u_{LS}(t)$ represents a more suitable choice than $u(t)$ in equation (\ref{eq:MTH}) since it is assumed that fluctuations in the large scale velocity, $u_{LS}(t)$, drive the variations in the friction velocity affecting the behaviour of small scales (see relation (\ref{eq:utau_uLS}) and accompanying discussion). In this regard, the relation (\ref{eq:MTH}) leads to scaling arguments that are in good agreement with the quasi-steady quasi-homogeneous theory, as it will be discussed in \S~\ref{subsec:res_u_QS}.

	One additional feature emerging from \figurename~\ref{fig:res_knp_U} concerns the reversal in the modulation mechanism from the wall proximity to the outer flow. In fact, while \figurename~\ref{fig:res_knp_U} shows a continuous decrease of $K_{np}$ for $y^+>10$ and $y/h \lesssim 0.5$, an almost absence of FM was previously found in the log-region by means of other techniques (where a reversal of the FM was only detected in the proximity of the end of the boundary layer)~\citep{ganapat2012amplitude, baars2015wavelet}. We point out that the $K_{np}$ behaviour in \figurename~\ref{fig:res_knp_U} resembles the decreasing behaviour of the AM parameters~\citep[e.g., see ][]{mathis2009Large, mathis2009comparison}, with a reversal of $K_{np}$ from the near-wall (where $K_{np}>1$) towards the outer region (where $K_{np}<1$). In principle, if the small scales are both amplitude- and frequency-modulated, one could expect that -- following the Newtonian principle that \textit{to the same natural effects we must, as far as possible, assign the same causes} -- a similar underlying mechanism is at play for both amplitude and frequency modulation. This can justify the similarity between the $K_{np}(y^+)$ behaviour -- that quantifies FM -- with the widely reported behaviour of AM parameters, either in internal or external wall-bounded turbulent flows. In other words, both AM and FM result from a common underlying phenomenon, for which both amplitude and frequency of small scales are concurrently affected by negative or positive large-scale fluctuations at different wall-normal locations.		
	 	\\Specifically, concerning the reversal coordinate (i.e., the $y^+$-location where $K_{np}$ switches from $K_{np}>1$ to $K_{np}<1$), for amplitude modulation the reversal typically occurs in the middle of the log-region, $y^+\approx 3.9 \Rey_\tau^{1/2}$~\citep{mathis2009Large, mathis2009comparison, ganapat2012amplitude, baars2015wavelet}. For frequency modulation, instead, it is still not clear whether a modulation reversal does occur in the log-region (as shown in \figurename~\ref{fig:res_knp_U}) or it is only limited to the wake region~\citep{baars2015wavelet}. From \figurename~\ref{fig:res_knp_U} we can conclude that FM mechanism is indeed limited to a near-wall region up to approximatively $y^+= 100$, consistently with the analysis by~\citet{ganapat2012amplitude}. Nevertheless, \figurename~\ref{fig:res_knp_U} also shows that the reversal $y^+$ location increases with the Reynolds number. In fact, we find that -- when spatial-series are focused both from channel DNSs and time-series by the MTH -- the reversal coordinates (i.e., $y^+\approx 35, 75, 145$ for the channel and boundary layer flows at $\Rey_\tau=1000, 5186, 14750$, respectively) scale as $y^+\approx 1.15\Rey_\tau^{1/2}$ (the power-law fit gives an exponent of $0.5$ with an $R^2=0.985$). This scaling has the same functional relationship found for amplitude modulation, i.e., $y^+\approx 3.9 \Rey_\tau^{1/2}$, but with a different proportionality constant. In particular, the $\Rey_\tau^{1/2}$-trend is reminiscent of the scaling of the outer peak position as a function of Reynolds number~\citep{mathis2009Large}, thus strengthening the underlying connection of FM with the change in large scale features.

		Another notable aspect discernible in \figurename~\ref{fig:res_knp_U} is the V-like shape of $K_{np}$ in the outer region of the channel flow, i.e. around $y/h\approx 0.5$ ($y^+\approx 2500$). In literature, the increase -- giving the V-like shape -- of amplitude modulation (AM) of small scales (which are representative of fine-scale turbulent motion) can be observed close to the channel center, e.g., in~\citet{chung2010Large} (see Figure 4 therein) for the streamwise velocity, $u$, or in~\citet{yao2018amplitude} (see Figure 3c therein) for the $v$ and $w$ components. However, to the best of our knowledge, this peculiar increase of the modulation parameter in the channel flow has not been explicitly discussed so far. Here we propose an interpretation based on the insights gained from turbulent boundary layers.
		\\Previous analyses, in fact, have highlighted that the preferential arrangement of the small scales in the wake region of turbulent boundary layers is mainly affected by intermittency, namely the presence of bulges of turbulent and non-turbulent flow~\citep{baars2015wavelet, baars2017reynolds}. However, \figurename~\ref{fig:res_knp_U}(a) shows a similar V-like behaviour -- as for previous results on AM -- also in the outer region of the channel flows, despite the absence of a turbulent/non-turbulent region in the channel flow (that is an internal flow). In particular, the V-like shape for the channel at $\Rey_\tau=1000$ -- although less evident than at $\Rey_\tau\approx 5200$ -- consistently occurs at $y/h\approx 0.5$ as highlighted in the inset of \figurename~\ref{fig:res_knp_U}(a).	Here we suggest that -- similarly to the effect of intermittency in boundary layer flows -- the preferential arrangement in the proximity of the channel centreline could be affected by an alternating occurrence of high- and low-rotational fluid motion above the head of large or very-large scale structures. This phenomenon would lead to the increase of both the AM and FM parameters toward the channel centreline. Although the clarification of this issue goes beyond the aim of this work, we do believe it deserves future investigations, being the channel flow setup much less considered for scale interaction analyses than turbulent boundary layers (for which high $\Rey_\tau$ data are much more available from experimental measurements).

		With the aim to ensure that the behaviour of $K_{np}(y^+)$ described so far is the result of an intrinsic flow phenomenon rather than an artefact due to the network representation, in \figurename~\ref{fig:res_knp_U} we also show the results for random-phase signals. Through a randomization of the phase of velocity Fourier coefficients, the energy spectral density and the turbulence intensity do not change, but any phase information is lost. Hence, following~\citet{mathis2009Large}, first the signals of $u$ are phase-scrambled, then the large scale component, $u_{LS}$, is extracted from the new random-phase signal (the amplitude spectrum is not changed), and eventually the degree is calculated from the full random-phase signal and conditioned to the sign of the random-phase $u_{LS}$. \figurename~\ref{fig:res_knp_U} shows that, both in the channel flow and boundary layer setups, the characteristic behaviour of $K_{np}$ for turbulent signals (black curves) disappears for random-phase velocity signals (blue curves). As previously reported \citep[e.g., see][]{chung2010Large}, phase relationships between large and small scales play an important role in the characterization of scale interaction, specifically on scale modulation; thus, if any realistic phase information is lost, modulation effects disappear as well.
		
	The NVG-based approach demonstrates to be reliable in capturing FM in turbulent velocity signals, and sensitive to phase-randomization. Moreover, the behaviour of $K_{np}(y^+)$ for both the channel and boundary layer is found to be robust under different cut-off wavelengths (used to extract $u_{LS}$ from the full signal $u$), as discussed in Appendix~\ref{app_sensit}. Further insights into the frequency modulation of the streamwise velocity will be presented in the next Section~\ref{subsec:res_u_QS} focusing on the near-wall region, i.e., where large-to-small scale modulation does essentially take place.

	\subsection{Scale interaction in the near-wall region}\label{subsec:res_u_QS}
	
		The presence of a near-wall modulation, whose intensity increases with the Reynolds number, has posed a challenge to the classical view on the universality of near-wall turbulence, i.e., the independence of the near-wall statistics (scaled in wall units) to the Reynolds number when this is sufficiently large. Since large scale structures affect the behaviour of the wall shear stress~\citep{mathis2013estimating}, the classical universality hypothesis has been recently replaced with the hypothesis that statistics have to be normalized by considering the large scale skin friction, $\tau_{LS}$, rather than the mean skin friction, $\tau_w$~\citep{zhang2016quasisteady, chernyshenko2020extension}. This hypothesis is referred to as quasi-steady quasi-homogeneous (QSQH) hypothesis, since the temporal and spatial variations of the large scale structures are much slower than variations of the near-wall turbulence~\citep{zhang2016quasisteady}. 
		
		The aim of this section is (i) to provide the proportionality relationships between $u_{LS}$ and the (temporal or spatial) frequency of the small scales as expected from the QSQH hypothesis, and (ii) to test the validity of such relationships by means of the network degree centrality. In particular, we will focus on velocity signals extracted at $y^+=15$ that corresponds to the $y^+$ value of maximum $K_{np}$ in \figurename~\ref{fig:res_knp_U}, thus being a representative wall-normal coordinate of the near-wall region. This choice is also related to the fact that, how evidenced by~\citet{zhang2016quasisteady} and~\citet{agostini2019departure}, the validity of the QSQH hypothesis is found to be rather accurate only in a narrow region close to the wall that is $y^+<70-80$.
		
		According to the QSQH hypothesis, variations in the large scale velocity, $u_{LS}$, induce proportional variations in the large scale skin friction, $\tau_{LS}$, namely $(\tau_w+\tau_{LS}) \propto (U+u_{LS})$, where $U$ is the local mean velocity. Since, by definition, $U_\tau=\sqrt{\tau_w/\rho}$ (where $\rho$ is the fluid density), the effect of $u_{LS}$ on $\tau_{LS}$ can be stated in terms of velocities as
		\begin{equation}\label{eq:utau_uLS}
			(U_\tau+u_{\tau,LS})\propto  \sqrt{(\tau_w+\tau_{LS})} \propto \sqrt{(U+u_{LS})}, 
		\end{equation}
		\noindent where $u_{\tau,LS}$ is the fluctuating (i.e., zero-mean) large scale component of the friction velocity~\citep{baars2017reynolds}. Due to near-wall modulation, a quasi-steady or quasi-homogeneous variation of the friction velocity affects also the (amplitude and) frequency of the small scales. Specifically, $u_{LS}>0$ events induce $u_{\tau,LS}>0$ (see relation~(\ref{eq:utau_uLS})) that, in turn, produces an increase of the small scale instantaneous spatial or temporal frequency according to the FM mechanism; \textit{vice versa} for $u_{LS}<0$~\citep{baars2017reynolds}.

		In the near-wall region, the spatial scales are supposed to have a constant characteristic length when normalized in wall-units (e.g., see the inner spectral peak for $\lambda_x^+=\lambda_x U_\tau/\nu=const.\approx 1000$ in \figurename~\ref{fig:spectra_deg}(a)). Therefore, spatial scales are related to $u_{\tau,LS}$ variations as $\lambda_x (U_\tau+u_{\tau,LS})=const.$, namely $\lambda_x\propto 1/(U_\tau+u_{\tau,LS})$. Since spatial frequency (i.e., wavenumber), $\kappa_x$, is related to spatial scales as $\kappa_x\propto \lambda_x^{-1}$, by using the~(\ref{eq:utau_uLS}) we obtain the following scaling relations
		\begin{equation}\label{eq:kappa_scaling}
			\lambda_x\propto {(U+u_{LS})}^{-1/2}, \quad \kappa_x\propto {(U+u_{LS})}^{1/2}.
		\end{equation}
		\noindent In the case of time-series, the temporal frequency, $f$, is related to spatial scales as $f=U_c/\lambda_x$, where $U_c$ is the convection velocity. Assuming that $U_c$ scales in wall units~\citep{baars2017reynolds}, namely $U_c\propto (U_\tau+u_{\tau,LS})$, temporal frequency is eventually expected to scale as 
		\begin{equation}\label{eq:f_scaling}
			f\propto (U_\tau+u_{\tau,LS})^2\propto (U+u_{LS}).
		\end{equation}
		\noindent Therefore, the relations~(\ref{eq:kappa_scaling}) and~(\ref{eq:f_scaling}) represent the expected scaling of spatial and temporal frequency, respectively, according the QSQH hypothesis.

		As discussed in Section~\ref{subsec:degree_FM}, the degree centrality represents a measure of the instantaneous wavelength or a temporal period. Therefore, it is expected that the degree, $k$, scales as $k\propto \lambda_x$ for the channel flow (in which spatial-series are mapped into NVGs), and $k\propto 1/f$ for the boundary layer (in which time-series are analysed). If the degree is indeed an effective parameter to quantify FM, $k$ should then be proportional to $(U+u_{LS})^{\beta_{u,x}}$ in the channel flow and $(U+u_{LS})^{\beta_{u,f}}$ in the boundary layer. Following the aforementioned scaling arguments (i.e., relations~(\ref{eq:kappa_scaling}) and~(\ref{eq:f_scaling})), the two exponents that verify the QSQH hypothesis should be equal to $\beta_{u,x}=-0.5$ and $\beta_{u,f}=-1$.

		\begin{figure}
		  \centerline{\includegraphics{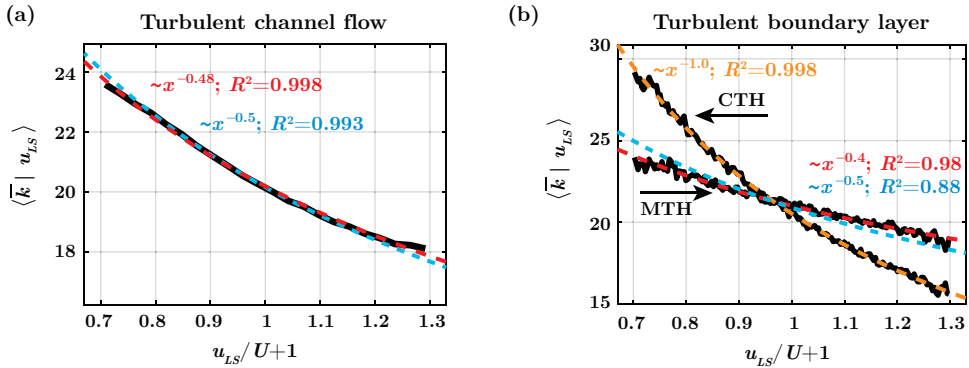}}
		  \caption{Average degree conditioned to the $u_{LS}$ values as a function of the normalized $u_{LS}$ deviation, for $u$ signals at $y^+\approx 10$ in (a) the turbulent channel flow and (b) the turbulent boundary layer (angular brackets indicate the average over time and homogeneous directions). In panel (b), the scaling for spatial-series obtained through the classical (CTH) and modified (MTH) Taylor's hypothesis are shown as black lines. The power-law fitting curves are shown as dashed lines, together with the exponent of the fitting and the coefficient of determination, $R^2$, for both setups. Light-blue dashed lines correspond to the expected scaling trends for spatial data. The intervals of $u_{LS}/U+1$ in abscissa cover, for each setup, a range from the $5$th to the $95$th percentile of all $u_{LS}$ at the selected vertical coordinate $y^+\approx 10$.} \label{fig:res_QS_u}
		\end{figure}

		To test the $\beta_{u,x}$ and $\beta_{u,f}$ scaling, we conditionally averaged the degree centrality values (computed from NVGs of the full velocity signals $u(x)$) to the $u_{LS}$ values at $y^+\approx 10$. In particular, $u_{LS}$ values were firstly divided into uniformly-binned intervals in the range $\textrm{min}[u_{LS}]-\textrm{max}[u_{LS}]$. Then, for each visibility network (i.e., each signal), nodes $i$ for which $u_{LS}(i)$ belongs to a specific bin were selected, and the corresponding degree values, $k_i$, were averaged for that specific bin. By extending the averages to all $u_{LS}$ bins, the conditional average, $(\overline{k}|u_{LS})$, is obtained, where the overbar indicates an average over a set of nodes. When plotting $(\overline{k}|u_{LS})$, the $u_{LS}$ value representative of each bin is chosen as the middle value of the bin.

		The behaviour of $(\overline{k}|u_{LS})$ as a function of $u_{LS}$ reveals the scaling between degree-based frequency variations and large scale velocity variations. Such conditional degree averages are shown in \figurename~\ref{fig:res_QS_u} for the channel flow at $\Rey_\tau\approx 5200$ (\figurename~\ref{fig:res_QS_u}(a)) and the boundary layer (\figurename~\ref{fig:res_QS_u}(b)), as a function of $(u_{LS}/U+1)$, where $U=U(y^+=10)$ is constant and $(u_{LS}/U+1)$ values equal to $1$ correspond to large scale zero-crossing points ($u_{LS}=0$). We find a scaling of the conditioned degree which follows a power-law with best-fit exponent $\beta_{u,x}=-0.48$ for the channel flow (spatial data) and $\beta_{u,f}=-1$ for the boundary layer when local mean velocity is used in the Taylor's hypothesis (CTH). These exponent values are in excellent agreement with the expected values of $-0.5$ and $-1$. We recall that, since for the CTH case the convection velocity is constant and equal to the local mean velocity $U$, the scaling exponent obtained in the CTH case is representative of a (temporal) frequency, thus obtaining $\beta_{u,f}=-1$. On the other hand, when the modified Taylor's hypothesis (MTH) is employed (equation (\ref{eq:MTH})), the structure of the spatial-series (obtained from the corresponding time-series) significantly changes and scaling arguments are therefore congruent with DNS spatial-series. Accordingly, the $(\overline{k}|u_{LS})$ scaling for the MTH case in \figurename~\ref{fig:res_QS_u}(b) produces an exponent which is close to $-0.5$, as expected from spatial-series. Finally, we mention that an exponent $\beta_{u,x}=-0.5$ is found for the channel flow at $\Rey_\tau= 1000$ with an $R^2\approx 0.92$ when cut-off filter is set to $\lambda_{x,c}^+=2500$, while for larger $\lambda_{x,c}^+$ values a poorer fitting is observed, likely due to the limited scale separation for this setup.

		While conditional averages were performed here by using uniformly-binned intervals of $u_{LS}$,~\citet{baars2017reynolds} -- by adopting a variable-interval scheme for conditional averages -- reported a scaling of approximatively $0.8$ (instead of $1$) for frequency in turbulent boundary layers over a wide range of $\Rey_\tau$. They indicated that the discrepancy in the expected exponent might be caused by an inaccurate assumption that small scales are convected at a fixed inner-scaled velocity. However, here we show that the expected scaling for $f$ is still obtained by assuming $U_c\propto (U_\tau+u_{\tau,LS})$, suggesting that the discrepancy in the fitting in~\citet{baars2017reynolds} might be related to different methodological arguments.

		The relevance of the scaling shown in \figurename~\ref{fig:res_QS_u} is twofold. From one hand, it demonstrates that -- similarly to the near-wall AM~\citep{baars2017reynolds} -- the near-wall FM agrees with the quasi-steady quasi-homogeneous hypothesis. On the other hand, the outcomes of \figurename~\ref{fig:res_QS_u} further validate the capability of the visibility-based approach -- relying on the degree centrality -- to capture FM in wall-bounded turbulence, as well as the validity of the modified Taylor's hypothesis (\ref{eq:MTH}) in converting time-series into spatial-series.

	\subsection{Analysis of the spanwise and wall-normal velocity components}\label{subsec:res_vw}

\begin{figure}
		  \centerline{\includegraphics{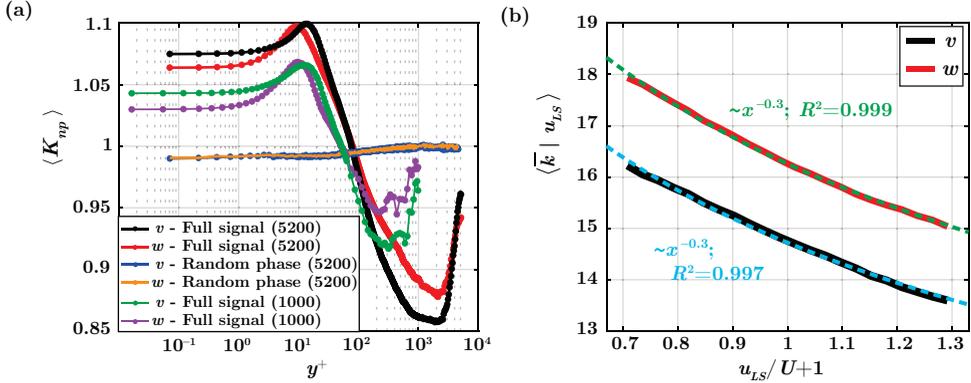}}
		  \caption{(a) $K_{np}$ ratio as a function $y^+$ for the wall-normal and spanwise velocity components, $v$ and $w$, extracted from the two channel flow DNSs, together with the corresponding $K_{np}$ values for random-phase signals. The respective Reynolds number value of the DNS is reported within brackets in the legend. Angular brackets indicate averaging over time and spanwise direction. (b) Average degree conditioned to the $u_{LS}$ values as a function of the normalized $u_{LS}$ deviation, for $v$ and $w$ signals at $y^+\approx 10$ in the turbulent channel flow at $\Rey_\tau\approx 5200$. The power-law fitting curves are shown as cyan and green dashed lines for the $v$ and $w$ cases, respectively, together with the exponents of the fitting and the coefficients of determination, $R^2$.} \label{fig:res_vw}
		\end{figure}

		The application of the NVG approach to spatial-series of wall-normal, $v(x)$, and spanwise, $w(x)$, velocity from the DNS of the turbulent channel flows is here reported. \figurename~\ref{fig:res_vw}(a) shows the ratio $K_{np}$ for the $v$ (black and green curves) and $w$ (red and purple curves) components as a function of $y^+$ for the two channel flows at different $\Rey_\tau$. $K_{np}>1$ is found in the near-wall region indicating a positive frequency modulation of the large scales on the small scales of $v$ and $w$. This result is consistent with amplitude modulation investigations, which show a similar modulating effect of the large scale motion on the small scales of the three velocity components~\citep{hutchins2007large, talluru2014amplitude, yao2018amplitude, wu2019modelling}. In particular, the trends shown in \figurename~\ref{fig:res_vw}(a) are similar to that reported in \figurename~\ref{fig:res_knp_U}(a) for the $u$ signals, although lower $K_{np}$ values are obtained from $v$ and $w$. Moreover, as for $u$, the FM of small scales of $v$ and $w$ is weaker at lower Reynolds number, since smaller $K_{np}$ values are observed in the near-wall region for $\Rey_\tau=1000$.
		\\Furthermore, similarly to the $u$ component, an almost constant $K_{np}\approx 1$ behaviour is found for network built from the random-phase $v$ and $w$ signals (see orange and blue curves in \figurename~\ref{fig:res_vw}(a) referring to $\Rey_\tau\approx 5200$ as a representative case), confirming the ability of the degree to capture phase information from the (full) signal. It should be noted that, for the sake of consistency, a unique phase shuffling was performed in this case for the three velocity components. As the degree from the $v$ and $w$ signals are conditionally averaged on $u_{LS}$, the phase of the streamwise velocity signal was extracted and randomly shuffled, so that the random-phase $u_{LS}$, $v$ and $w$ signals were obtained via the respective (non-shuffled) amplitudes but with the same random phases.

		Although spectral peak separation in the spectrograms of the transversal velocity components is less evident than for the streamwise velocity, the generation and amplification of small scale motions (i.e., fine scale vortices) of all the three velocity components is strongly connected with large scale events~\citep{hutchins2007large}. Within this perspective, the wall-normal and spanwise velocities are expected to be modulated in the near-wall region by following the QSQH hypothesis in a similar way as the streamwise component, $u$. However, while results for AM of the three velocity components~\citep{talluru2014amplitude, agostini2019departure, chernyshenko2020extension} and scaling arguments for the AM and FM of the $u$ component~\citep{baars2017reynolds} have been provided, as far as we know, similar scaling arguments (as in~\citet{baars2017reynolds}, figure 9 therein) for $v$ and $w$ have not been pursued for FM to date.
		
		In analogy with the modulation of the $u$ component, the conditional average degree, $(\overline{k}|u_{LS})$, is evaluated as a function of $(u_{LS}/U+1)$ at $y^+\approx 10$ for NVGs built from $v(x)$ and $w(x)$ signals. The conditional average degree and the corresponding fitting are shown in \figurename~\ref{fig:res_vw}(b) for the $\Rey_\tau\approx 5200$ setup and confirm the power-law modulation effect of the large scales in the near-wall region even for the other velocity components, namely $k\propto (U+u_{LS})^{\beta_{v,x}}$ and $k\propto (U+u_{LS})^{\beta_{w,x}}$. However, while for the $u$ component the exponent of the power-law was $\beta_{u,x}\approx -0.5$, a weaker scale interaction effect is found for the $v$ and $w$ components being $\beta_{v,x}\approx -0.3$ and $\beta_{w,x}\approx -0.3$, which are both smaller (in modulus) than $\beta_{u,x}$. This outcome is also consistent with the smaller $K_{np}$ values for $v$ and $w$ (see \figurename~\ref{fig:res_vw}(a)) than for $u$ (see \figurename~\ref{fig:res_knp_U}(a)), indicating a weaker FM in the near-wall region for the transversal velocity components. 
		\\The power-law relationships found for $u$, $v$ and $w$ suggest that – although the intensity of modulation is different for each velocity component -- the response of the small-scales of $v$ and $w$ exhibits a functional relation qualitatively analogous to the response of $u$. In general, there could be several factors playing a role in the scale-interaction mechanisms (such as the direction of the large-scale motions as discussed by~\citet{chernyshenko2020extension}), but we can conclude that the QSQH hypothesis is valid for all velocity components, although a more refined description is required for the transversal components, $v$ and $w$.

		The results shown in this section reveal that the three velocity components are all affected by a large scale FM in the near-wall region, where an increase of the local (spatial) frequency is observed under $u_{\tau,LS}>0$ periods induced by positive $u_{LS}$ events. In particular, we provided novel insights on FM for the $v$ and $w$ components -- which have been investigated less than $u$ -- in terms of FM intensity for spatial-series (\figurename~\ref{fig:res_vw}(a)), and scaling arguments on the QSQH hypothesis (\figurename~\ref{fig:res_vw}(b)).

	\subsection{Time and space shifting in FM}\label{subsec:res_shifting}
	
		To conclude our analysis, we provide results on the investigation of time- and space-shifted FM, as quantified by $K_{np}$. We recall that a lead of the small scale amplitude was found with respect to the large scales in the near-wall region of turbulent boundary layers, while a small scale lag is found above the reversal coordinate~\citep{bandyopadhyay1984coupling, guala2011interactions}. Concerning FM, a lead of the small scale frequency with respect to large scales was found in the near-wall region but, differently from AM, scattered behaviours were found far from the wall~\citep{ganapat2012amplitude, baars2015wavelet}.

		\begin{figure}
		  \centerline{\includegraphics{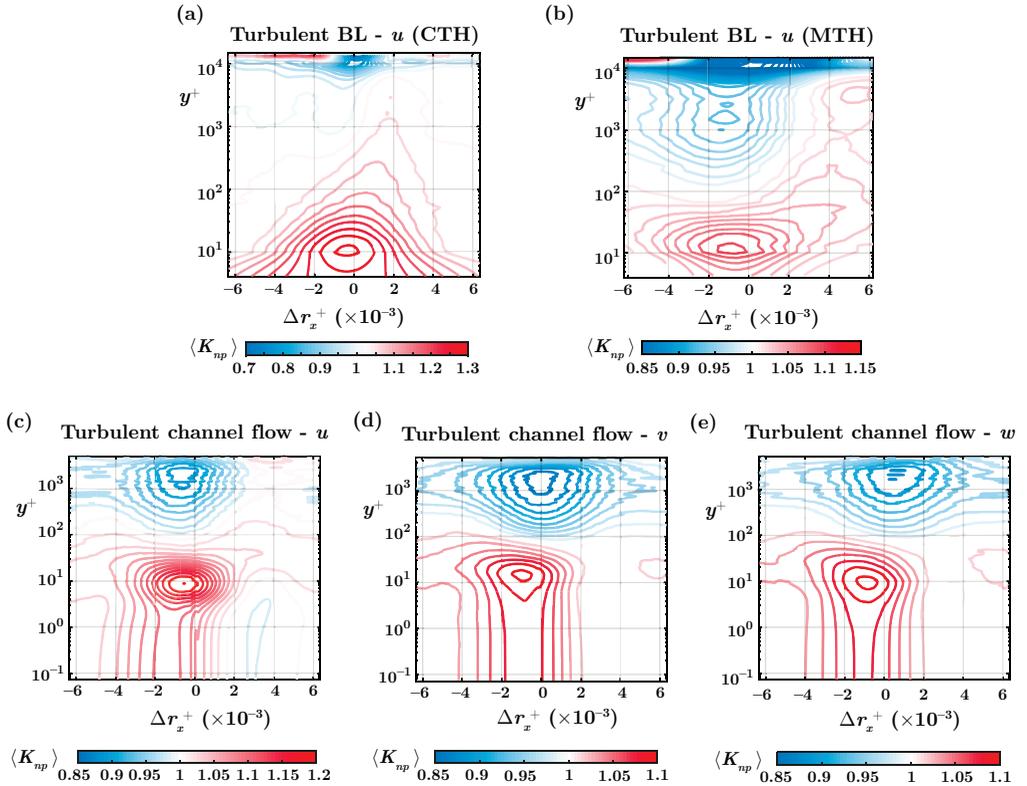}}
		  \caption{Contour plot of the $K_{np}$ ratio as a function of the wall-normal coordinate, $y^+$, and the temporal or spatial shifting, $\Delta t^+$ or $\Delta r_x^+$, respectively. Shifting for the turbulent boundary layer are reported in panels (a) and (b) whether the classical or modified Taylor's hypothesis is used, respectively. Spatial shifting for the turbulent channel flow at $\Rey_\tau\approx 5200$ is shown in panels (c-e) for the three velocity components. Iso-level contours are displayed by using a level-step equal to $0.03$ in panel (a), $0.01$ in panel (b) and $0.015$ in panels (c)-(e).} \label{fig:res_shifts}
		\end{figure}
		
		To address this issue, we show in \figurename~\ref{fig:res_shifts} the conditionally average degree, $K_{np}$, as a function of $y^+$ and the spatial delay, $\Delta r_x^+$, where for time-series it holds the Taylor's hypothesis $\Delta r_x^+=-U_c\Delta t^+$ (the minus sign highlights the opposite direction of reference systems between fixed-point time-series and spatial-series). Therefore, the formulation of $K_{np}$ reported in (\ref{eq:dndp}) is extended to account for spatial shifting, $\Delta r_x^+$, as $K_{np}(\Delta r_x)=K_n(\Delta r_x)/K_p(\Delta r_x)$, with

	\begin{equation}\label{eq:dndp_shift}
	\left\{ \begin{array}{ll}
			K_n(\Delta r_x)=\frac{1}{N_{neg}(\Delta r_x)}\sum_{j=1}^N {\left(k(x_j-\Delta r_x)|u_{LS}(x_j)<0\right)},\\[10pt]
			K_p(\Delta r_x)=\frac{1}{N_{pos}(\Delta r_x)}\sum_{j=1}^N{\left(k(x_j-\Delta r_x)|u_{LS}(x_j)>0\right)}.
		\end{array}\right.
	\end{equation}
	
	\noindent Positive or negative $\Delta r_x^+$ values indicate in (\ref{eq:dndp_shift}) a lag or lead, respectively, of the degree, $k$, with respect to $u_{LS}$ in the conditional averages of equation~(\ref{eq:dndp}) (the results in \figurename~\ref{fig:res_knp_U} correspond to $\Delta r_x^+=\Delta t^+=0$). If classical Taylor's hypothesis (CTH) is employed for turbulent boundary layer time-series (\figurename~\ref{fig:res_shifts}(a)), a slight lead of $K_{np}$ with respect to $u_{LS}$ (i.e., high $K_{np}$ values are at $\Delta r_x^+<0$ but close to $\Delta r_x^+=0$) is observed for $y^+\lesssim 15$. However, a more substantial lead is observed for larger $y^+$ coordinates in the near-wall region up to $y^+\approx 100$, in agreement with previous analyses~\citep{baars2015wavelet}, while a lag of $K_{np}$ with respect to large scales is detected for $y^+\gtrsim 100$. A clearer picture is obtained when the modified Taylor's hypothesis (MTH) is exploited (\figurename~\ref{fig:res_shifts}(b)). Significant lead of $K_{np}$ with respect to $u_{LS}$ is found in the whole near-wall region (including wall proximity, $y^+\lesssim 15$), while the lag for $y^+\gtrsim 100$ is less evident and a lead is recovered for larger $y^+$ values (see blue contours in \figurename~\ref{fig:res_shifts}(b)). Eventually, no clear patterns are observed in the intermittent regions ($y^+\gtrsim 5\times 10^3$).

		The space-shifted $K_{np}$ values in the turbulent channel flow at $\Rey_\tau\approx 5200$ for the three velocity components are displayed in \figurename~\ref{fig:res_shifts}(c-e). Likewise the turbulent boundary layer, a lead of $K_{np}$ with respect to $u_{LS}$ is found for $y^+<100$, as highlighted by high $K_{np}$ values for $\Delta r_x^+<0$. Differences among the $u$, $v$ and $w$ components are detected in proximity of the channel centreline, where $K_{np}$ appears to lead, be in-phase, and slightly lag $u_{LS}$ for the $u$ (\figurename~\ref{fig:res_shifts}(c)), $v$ (\figurename~\ref{fig:res_shifts}(d)), and $w$ component (\figurename~\ref{fig:res_shifts}(e)), respectively. It should be noted that the $\Rey_\tau\approx 5200$ is here used as a representative setup for spatial data, and results for the lower $\Rey_\tau$ channel flow are in agreement with results in \figurename~\ref{fig:res_shifts} so they are not shown for the sake of conciseness.
		\\In particular, it is worth highlighting that the maximum $K_{np}$ values in the near-wall region are found at $\Delta r_x^+\approx 1000$ for the streamwise velocity in the turbulent boundary layer when the MTH is employed (\figurename~\ref{fig:res_shifts}(b)), as well as for all the velocity components in the turbulent channel flow (\figurename~\ref{fig:res_shifts}(c-e)). The value $\Delta r_x^+\approx 1000$ is in very good agreement with the characteristic length scale in the near-wall region, being $\lambda_x^+=O(10^3)$ (see inner spectral peak in \figurename~\ref{fig:spectra_deg}(a)). The equivalent time-scale is $\Delta t^+=\Delta r_x^+/U_c^+\approx 100$ (being $U_c^+\approx 10$ in the buffer layer and viscous sublayer), which is the characteristic turnover time of the near-wall cycle. As small scales are supposed to be actively modulated by large-scales, the time taken for this process to be completed is therefore equivalent to the time scale of the near-wall cycle~\citep{ganapat2012amplitude}. 		
		
		The results shown in \figurename~\ref{fig:res_shifts}(a-c) for the streamwise velocity reveal that different convection velocities indeed play a significant role in the FM dynamics, not only in terms of overestimation (as highlighted in \figurename~\ref{fig:res_knp_U}(b)), but also in terms of spatial delay that -- in the near-wall region -- is strongly related to the near-wall cycle. Therefore, NVG reveals again to be a reliable approach for quantifying FM even in presence of a temporal or spatial shifting. Finally, we note that an important issue about scale interaction is whether large scales actually cause an increase or decrease of small scale activity, as the parameters used so far to quantify AM and FM only show there is a relation (e.g., a correlation) between large scales and small scales. Although definite answers to this issue are not still available, our detection of the presence of a significant temporal or spatial delay close to the characteristic time or length scale of the near-wall cycle, in conjunction with the arguments leading to relation~(\ref{eq:utau_uLS}), could provide supporting clues that a causation process is at play. In fact, fluctuations in the large scale component of the wall shear stress -- which affect the small scales behaviour -- appear to be directly caused by the outer large scale structures rather than being the feature of near-wall processes~\citep{zhang2016quasisteady}.

\section{Discussion}\label{sec:discuss}

	In this work, the natural visibility graph was used to study 1D spatial-series and time-series from two turbulent flow configurations, but some generalizations can be carried out. First, the geometrical criterion at the basis of the visibility algorithm can be extended to scalar fields of arbitrary dimension~\citep{lacasa2017visibility}. For instance,~\citet{tokami2020spatiotemporal} recently constructed a spatial visibility graph (employing a simplified version of the NVG called horizontal visibility graph as proposed by~\citet{luque2009horizontal}) from a 2D velocity field in a buoyancy-driven turbulent fire. Therefore, our approach could be extended to 2D velocity fields at fixed $y^+$ coordinates, thus concurrently taking into account the degree variations along the streamwise and spanwise directions.

	The results provided by natural visibility graphs, specifically about the degree centrality, necessarily depend on the signal resolution (either the sampling frequency or the grid size), which needs to be sufficiently high to capture the behaviour of small scales. However, if the temporal or spatial resolution is sufficient enough to capture the smallest significant features of the signal, the degree centrality tends to proportionally scale with the signal resolution as shown, e.g., in~\citet{iacobello2018visibility}. Nevertheless, an additional feature of visibility networks is the possibility to explicitly account for the spatial or temporal discretization. In fact, one can assign to each discrete observation, $i$, the corresponding signal spacing (e.g., $\Delta x_i$, $\Delta z_i$, $\Delta t_i$, etc.). In this way, each node $i$ of the network is representative of an interval centred in $i$, thus providing a continuous representation of the signal. As a result, a weighted network is obtained in which the relation~\ref{eq:degree} is reformulated as $\tilde{k}_i\equiv\sum_j{\Delta \chi_j  A_{i,j}}$, for a series sampling, $\Delta\chi_j$, where $\chi$ is independent variable (e.g., time). This generalization is particularly useful for non-uniformly sampled signals from experimental measurements, in which $\tilde{k}$ can be used in place of $k$, e.g., in the definition~(\ref{eq:dndp}).

	Finally, it is worth to observe that, so far, the visibility approach was presented as a convexity criterion (see inequality~(\ref{eq:visib})). In particular, the network degree was interpreted as a measure of the instantaneous period (quantified in terms of the local convexity of the signal), in analogy with the concept of instantaneous frequency based on the Hilbert transform (where the local properties of a series are emphasized by performing a convolution of the signal with the function $1/t$~\citep{huang1998empirical}). Nevertheless, the visibility algorithm can also be used as \textit{concavity} criterion by applying it to the opposite signal, $-s_i$, whose effect is to change the direction in the inequality~(\ref{eq:visib})~\citep{iacobello2019experimental}. The comparison of the network metrics extracted from $s_i$ and $-s_i$ allows one to characterize the peak-pit asymmetry of a signal, especially in real-world phenomena~\citep{hasson2018combinatorial}. Following this point of view, we evaluated -- for the sake of completeness -- the values of $K_{np}(y^+)$ by using the NVG as a concavity criterion for the streamwise velocity, and we found that the main features of the FM for full and random-phase signals are retained either when the information is only taken from the convexity or concavity criterion.

\section{Conclusions}\label{sec:conclusion}

	In this study, we propose a novel approach to investigate the frequency modulation (FM) mechanism in wall-bounded turbulence by means of the natural visibility graphs. Spatial-series and time-series of the velocity from two turbulent channel flows and a turbulent boundary layer, respectively, are mapped into visibility networks and the degree centrality is conditionally averaged to the sign of the large scale velocity to quantify FM. In particular, the versatility of visibility graphs to map either time- or spatial-series, let us exploit velocity spatial-fields from turbulent channel flows that have been much less investigated than turbulent boundary layers under the lens of frequency modulation.

	The overall results for the streamwise velocity indicate a frequency modulation mechanism occurring in the near-wall region with a peak of intensity in the buffer layer, in agreement with previous works. However, in contrast with previous observations on FM, we observe a reversal in the frequency modulation mechanism far from the wall similarly to what observed for amplitude modulation, in both channel and boundary layer flows. We argued that such similarity could stem from a common underlying phenomenon, for which both amplitude and frequency of small scales are concurrently affected by negative or positive large-scale fluctuations. Moreover, we observe that the reversal coordinate scales as ${\Rey_\tau}^{0.5}$, which is reminiscent of the scaling in the wall-normal position of the outer spectral peak.

	The effect of different convection velocities for the time-series analysis is also discussed. In particular, we modified the correction proposed by~\cite{yang2018implication} by accounting for only the large scale velocity component in the definition of the convection velocity. This choice is based on the rationale that variations in the large scale velocity induce variations in the wall shear stress, which in turn affect the behaviour of small scales. We detect an overprediction of frequency modulation when the local mean velocity is employed as convection velocity in the turbulent boundary layer, while such overprediction is compensated when the proposed modified Taylor's hypothesis is used. Moreover, scaling behaviours of the degree centrality as a function of the large scale velocity are found to be in very good agreement with the the quasi-steady quasi-homogeneous (QSQH) theory. In this regard, our correction of the Taylor's hypothesis provides reliable scaling trends as the large scale velocity is supposed to induce modulation of small scales through variations in the wall shear stress.

	Finally, the FM for the wall-normal and spanwise velocity components is analysed for the turbulent channel flows and FM scaling is discussed for the transversal velocities. We find a frequency modulation mechanism for the wall-normal and spanwise velocity components qualitatively similar to FM of the streamwise velocity. Specifically, a power-law scaling of the degree conditioned to the large scale velocity is found for the three velocity components, although smaller exponents are found for transversal velocities than for the streamwise velocity. Moreover, a delay-based analysis is carried out for the three velocity components in the channel flow and for streamwise velocity time-series in the boundary layer. We observe that small scales lead large scales in the near-wall region (in accordance with previous studies), but significant differences are found when the classical or modified Taylor's hypothesis is applied. Specifically, our modified Taylor's hypothesis provides results in agreement with spatial-series analysis, where the delay of maximum modulation corresponds to the characteristic length (or temporal) scale of the near-wall cycle.
	\\Furthermore, we emphasize here that, to the best of our knowledge, this is the first time that frequency modulation is thoroughly investigated for all the three velocity components, as previous works have been mainly focused on amplitude modulation.	The findings gained through the visibility networks of all the three velocities can then contribute to the development of a more general model of scale interaction, which accounts for the different modulating effect of the large scale on each velocity component.
	
	The visibility-based approach reveals to be robust in the quantification of FM with respect to AM (Appendix~\ref{app_synt}), and to different cut-off filtering sizes and high-frequency noise (Appendix~\ref{app_sensit}), as well as sensitive to a spectral phase randomization of the signals. The latter implies that the natural visibility graph is able to capture non-linearities in the signal, as linear effects are preserved during phase randomization (i.e., amplitude spectrum does not change) while non-linearities are lost through phase-shuffling. We stress that the visibility networks do not require any \textit{a priori} parameter, and are directly built from the full velocity signals (instead of the small scale component), being the network degree able to capture the signal structure at local scales. In this regard, although in this work a one-point analysis is carried out for simplicity, a two-point analysis (where the large scale signal is extracted at a fixed wall-normal coordinate) would reveal the full potential of visibility networks. In fact, when multiple synchronized signals are available at different wall-normal locations (e.g., from numerical simulations, hot-wire rakes or through particle image velocimetry), the large scale signal can be obtained only once at a fixed wall-normal location, as well as probes working on a smaller frequency range can be employed (being only low-frequencies necessary). The full velocity signals, instead, can be used without any filtering operation to capture the small scales frequency modulation at the remaining wall-normal locations.

	 In the wake of the recent successful applications of network science to the analysis of turbulent flows~\citep{murugesan2015combustion, taira2016network, schlueter2017coherent, krishnan2019emergence, iacobello2019lagrangian}, the proposed visibility-based approach candidates for being a parameter-free and robust tool for FM investigation.

\appendix

\section{Synthetic signals for visibility-based FM detection}\label{app_synt}

\begin{figure}
	  \centerline{\includegraphics{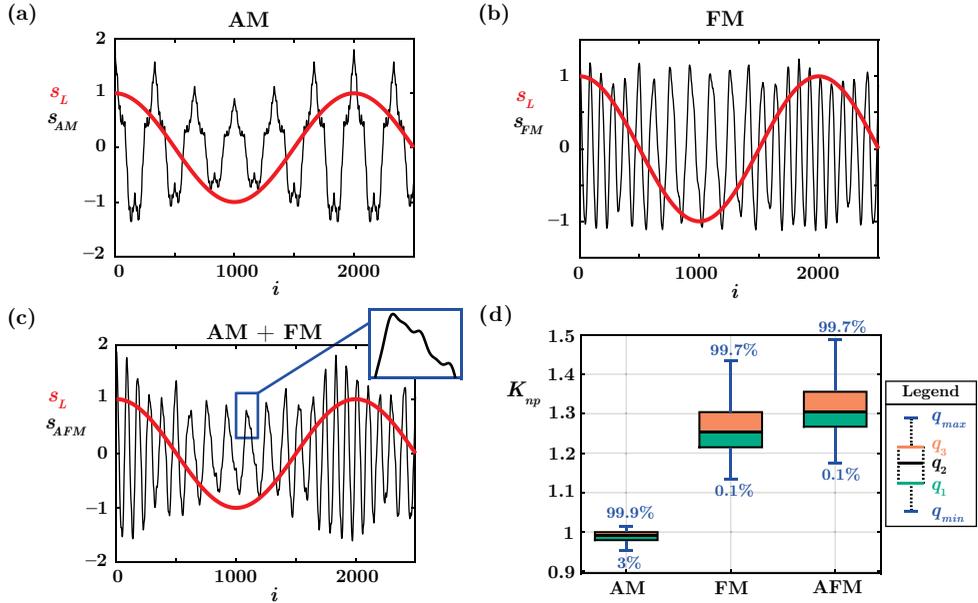}}
	  \caption{First 2500 time instants (out of $10^4$) of the three synthetic modulated signals (shown in black) and the modulating signal (shown in red) for: (a) amplitude modulated signal (AM); (b) frequency modulated signal (FM); (c) amplitude and frequency modulated signal (AFM). The inset show a zoom of the AFM modulated signal. (d) Results of the application of the NVG to the synthetic signals in panels (a-c). Values of $K_{np}$ are shown as box plots, where $q_1$, $q_2$ and $q_3$ are the 25th, median, and 75th percentiles, respectively, while $q_{min}=q_1-1.5(q_3-q_1)$ and $q_{max}=q_1+1.5(q_3-q_1)$ whose values are explicitly indicated at the tips of the whiskers as percentages.} \label{fig:eg_synt_FM}
	\end{figure}

	In this appendix we provide results of the application of the visibility-based approach to quantify frequency modulation from synthetic signals, which is a simple but representative benchmark of more complex signals such as from turbulent flow fields. Three configurations of modulation are here investigated, as shown in \figurename~\ref{fig:eg_synt_FM}(a-c), namely amplitude modulation (AM), frequency modulation (FM) and both amplitude and frequency modulation (AFM). In this way, we assess the effect of different modulations on the ratio $K_{np}$ and its ability to discern FM only.

	All the generated signals have length $N=10^4$ and sampling frequency $f_{samp}=4000$ Hz, which is chosen to be much larger than the characteristic frequencies of the modulated and modulating signals. The modulating (i.e., large scale) signal -- shown in red \figurename~\ref{fig:eg_synt_FM}(a-c) -- is given by the expression $s_L(t_i)=\cos(2\pi f_L t_i)/3$, where $t_i=(i-1)/f_{samp}$ is time, with $i=1,\dots,N$, and $f_L=2$ Hz is the frequency of the modulating signal. 
	\\The three modulated signals, $s_{AM}$, $s_{FM}$ and $s_{AFM}$, are constructed as high-frequency sinusoidal series, which emulate the small scale velocity component, modulated by $s_L$. A positive modulation is considered, namely, an increase of amplitude and/or frequency is induced for intervals of positive $s_L$ values, and \textit{vice versa} for negative $s_L$ values. This behaviour mimics the modulation close to the wall by large scales to small scales in wall-bounded turbulence. Specifically, the three modulated signals, shown as black lines in \figurename~\ref{fig:eg_synt_FM}(a-c), are given as follows:
	
	\begin{itemize}
		\item $s_{AM}=(\cos\left[2\pi f_H t_i\right]+s_R)(1+s_L)$, 
		\item $s_{FM}=\cos\left[2\pi f_H t_i + \varphi_L\right]+s_R$, 
		\item $s_{AFM}=s_{FM}(1+s_L)$,
	\end{itemize}
	
	\noindent where $f_H=12$ Hz is the (high) carrier-frequency of the modulated signals. The role of $(1+s_L)$ is to provide the amplitude modulation effect on $s_{AM}$ and $s_{AFM}$, while the role of $\varphi_L$ is to give a frequency modulated component to $s_{FM}$ and $s_{AFM}$. In particular, $\varphi_L$ is a time-varying phase depending on $s_L$ typically used to generate frequency-modulated signals~\citep{boashash2015time}, defined as $\varphi_L=2\pi f_\Delta I_m$, where $f_\Delta=11$ Hz is the \textit{frequency deviation} (i.e., the maximum frequency shift from $f_H$), and $I_m\equiv\sum_i{s_L(t_i)}/f_{samp}$. The frequency deviation, $f_\Delta$, is selected to be close to the value of the carrier frequency, $f_H$, in order to maximize the modulation effect on the signal.

	In each of the three modulated series, an additional signal, $s_R$, is also included. It is obtained as a sum of unmodulated cosine signals with randomly-varying amplitude, given by 
	
	\begin{equation} \label{eq:disturb}
		s_R(t_i)=\sum_{q=2}^5{\frac{r_A}{2^q}\cos\left[2\pi(2^q f_H)t_i\right]},
	\end{equation}		
	
	\noindent where $r_A$ is a random number extracted from a uniform distribution in the range $(0,1)$. The effect of $s_R$ in a modulated signal can be observed in the inset of \figurename~\ref{fig:eg_synt_FM}(c). The role of $s_R$ is to introduce -- similarly to turbulence velocity spectra -- additional high-frequency low-amplitude components, thus making the modulated (small scale) signal a broadband-like series.
		
	By generating several random amplitudes, $r_A$, in equation~\ref{eq:disturb}, an ensemble of $s_R$ series is obtained for each $r_A$. This leads to an ensemble of different modulated signals, $s_{AM}$, $s_{FM}$ and $s_{AFM}$, that are characterized by different $s_R$. Specifically, we generated $5\times 10^3$ values of $r_A$ for each of the three modulated signals. The values of the ratio $K_{np}$ (see equation~\ref{eq:dndp}) are then computed for each ensemble, by evaluating the degree on the NVGs built for the full signals, namely, $(s_{AM}+s_L)$, $(s_{FM}+s_L)$ and $(s_{AFM}+s_L)$.
	\\\figurename~\ref{fig:eg_synt_FM}(d) shows the values of $K_{np}$ for the three modulation configurations as box-plots, in which the most significant percentiles are highlighted. For the AM case, $K_{np}$ is concentrated around unity, with a median value that is very close to one, as expected since the main modulating effect is on amplitude. For the FM and AFM cases, instead, values of $K_{np}$ greater than one are consistently obtained (note the percentile values in \figurename~\ref{fig:eg_synt_FM}(d)), as a result of the positive frequency modulating effect of the large scale signal. In particular, it is worth noting that even when a signal is modulated both in amplitude and frequency, the ratio $K_{np}$ is able to emphasize the contribution of the FM. 
	
	The results shown in \figurename~\ref{fig:eg_synt_FM}(d) reveal that $K_{np}$ is an accurate parameter to quantify FM, since it consistently shows positive values under positive FM, and also a precise metric, since there is narrow spreading of the $K_{np}$ values around the median. The results shown in this appendix corroborate the ability and robustness of the proposed visibility-based approach -- relying on the conditioned degree centrality -- to capture frequency modulation, thus fostering its application as a tool to study scale-interaction in wall-bounded turbulence.

\begin{figure}
	  \centerline{\includegraphics{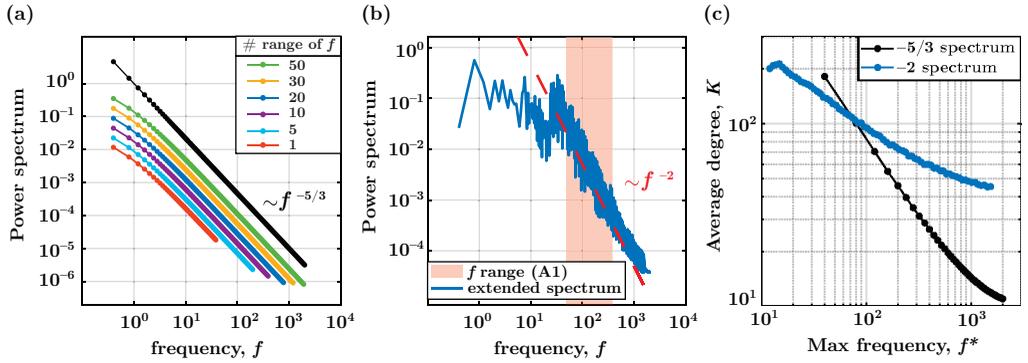}}
	  \caption{(a) Power spectra of synthetic signals following a $f^{-5/3}$ law. Signals are generated through inverse-Fourier transform of the power spectrum shown in black for different frequency ranges (i.e., different number of harmonics). Fifty ranges of $f$ are generated, and the spectra of the synthetic signals for some representative $f$ ranges are highlighted with different colours, as well as vertically shifted to enhance visualization. (b) Power spectrum of signals generated as per equation (\ref{eq:disturb}). The range of frequency considered for the equation (\ref{eq:disturb}) is highlighted as a shaded red region, and an $f^{-2}$ scaling is also shown. (c) Average degree centrality, $K$, for signals generated as in (a) and (b) for increasing maximum frequency, $f^*$.} \label{fig:synt_K_harmon}
	\end{figure}

	To conclude this section, we show the effect of higher frequency harmonics on the average degree, $K$, for synthetic signals. With this aim, we used two sets of synthetic signals according to their power spectrum scaling: (i) following a $-5/3$ spectrum, and (ii) a $-2$ spectrum. The former emulates turbulent signals in the inertial range, while the second refers to signals defined in equation (\ref{eq:disturb}). \figurename~\ref{fig:synt_K_harmon}(a-b) show the power spectra for both types of synthetic signals, while \figurename~\ref{fig:synt_K_harmon}(c) illustrates the behaviour of $K$ as a function of the maximum frequency considered, $f^*$. As discussed in \S~\ref{subsec:degree_features} referring to \figurename~\ref{fig:spectra_deg}(c) for turbulent series, $K$ decreases as the number of high-frequency harmonics increases. Moreover, the changes in $K$ are stronger for the synthetic signals following the $f^{-5/3}$ spectrum (\figurename~\ref{fig:synt_K_harmon}(a)) than for $f^{-2}$ spectrum (\figurename~\ref{fig:synt_K_harmon}(b)), because the energy content of small scales is larger in the former case being the exponent $-5/3$ lower (in modulus) than $-2$.

\section{Sensitivity analysis}\label{app_sensit}	

The aim of this section is to assess the robustness of the proposed NVG-based approach under different values of the spectral filtering wavelength and under high-frequency noise.

	We recall that the spectral filtering wavelength is used to extract the large scale component, $u_{LS}$, from the streamwise velocity signal, $u$. \citet{mathis2009Large} firstly reported a sensitivity analysis on the AM of streamwise velocity in a turbulent boundary layer. They showed that a decrease of the cut-off wavelength leads to a small increase of the AM below the reversal wall-normal coordinate (i.e., in the near-wall region), and a small decrease of AM above the reversal coordinate (i.e., far from the wall). The conclusion was that, despite the small variations due to different cut-off wavelengths, the general form of the AM parameter is retained. For this reason, subsequent works on AM and FM exploited the sensitivity analysis by~\citet{mathis2009Large} as a reference case to justify the choice of the cut-off wavelength.

\begin{figure}
	  \centerline{\includegraphics{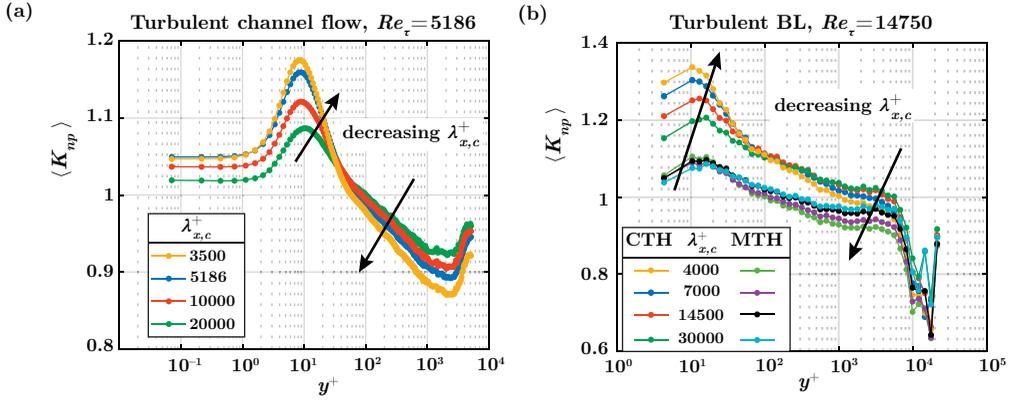}}
	  \caption{Effect of different cut-off wavelengths in the large scale conditional average degree ratio, $K_{np}(y^+)$, for streamwise velocity, $u$, extracted from (a) the channel flow DNS at $\Rey_\tau\approx 5200$ and (b) the boundary layer experiments. In panel (b), the cut-off effect for spatial series obtained from time-series via classical Taylor's hypothesis (CTH) -- namely using as convection velocity the local mean velocity -- and modified Taylor's hypothesis (MTH). Angular brackets indicate averaging over time and spanwise direction in (a) and over three different realizations in (b).} \label{fig:sensit_res}
	\end{figure}

	Here we perform a sensitivity analysis on the wall-normal behaviour of $K_{np}$ for the streamwise velocity, by changing the cut-off wavelength, $\lambda_{x,c}$. \figurename~\ref{fig:sensit_res} shows $K_{np}$ as a function of $y^+$ for four $\lambda_{x,c}$ values, in the turbulent channel flow at $\Rey_\tau\approx 5200$ (\figurename~\ref{fig:sensit_res}(a), $\lambda^+_{x,c}=5186$ in the main text) and the turbulent boundary layer (\figurename~\ref{fig:sensit_res}(b), $\lambda^+_{x,c}=7000$ in the main text). For the boundary layer, both the classical and modified Taylor's hypotheses are considered and labelled as CTH and MTH in the legend of \figurename~\ref{fig:sensit_res}(b). The nominal shape of $K_{np}$ as a function of $y^+$ is maintained both for the channel and boundary layers setups, and -- similarly to the analysis carried out by~\citet{mathis2009Large} -- a decrease in $\lambda_{x,c}$ leads to a reduction of $K_{np}$ below the reversal $y^+$ and a rise of $K_{np}$ above the reversal $y^+$. Specifically, variations of $K_{np}$ in the wall proximity are less evident for the boundary layer when the modified Taylor's hypothesis (MTH) is applied rather than when local mean velocity is considered as convection velocity (CTH).
	
	This sensitivity analysis confirms the robustness of the decoupling procedure to extract $u_{LS}$, which is employed to evaluate $K_{np}$ as metric for studying frequency modulation.
	
\begin{figure}
	  \centerline{\includegraphics{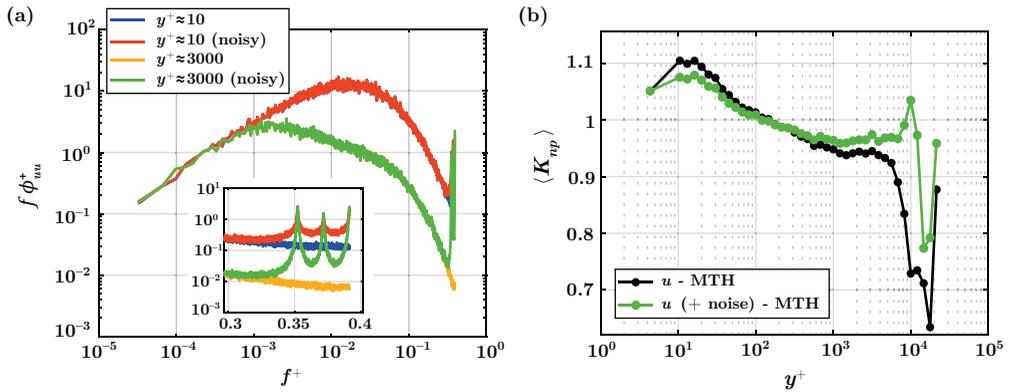}}
	  \caption{(a) Pre-multiplied energy spectrum of the streamwise velocity, $uu$, from the turbulent boundary layer at two representative $y^+$ locations with and without high-frequency noise. The inset shows a zoom at the highest frequencies. (b) $K_{np}$ as a function of $y^+$ for velocity signals without (black) and with (green) high-frequency noise (the MTH is used).} \label{fig:sensit_HFnoise}
	\end{figure}

	Finally, our method is tested under the presence  high-frequency noise in the velocity signals (as usually happens in experimental measurements). With this aim, we artificially added a high-frequency noise to experimental signals of the streamwise velocity (whose sampling frequency is $f_s=20000$ Hz) in the turbulent boundary layer. The noise signal is given by the sum of three harmonics with random phase and with frequencies equal to $0.5 f_s$, $0.475 f_s$ and $0.45 f_s$ (the maximum frequency included is $f_s/2$ as higher frequencies are not be captured in the amplitude spectrum), whose effects on spectra are displayed in \figurename~\ref{fig:sensit_HFnoise}(a). The corresponding values of $K_{np}$ are shown in \figurename~\ref{fig:sensit_HFnoise}(b), where we observe that the behaviour of $K_{np}$ is retained throughout the boundary layer except for the intermittency region where the noise intensity significantly affects the signal structure. Therefore, we conclude that -- although a pre-processing of the (experimental) data is always a good practice to avoid biased behaviours -- the NVG approach based on $K_{np}$ is sufficiently robust under high-frequency noise.
	
\backsection[Declaration of interests]{The authors report no conflict of interest.}	

\backsection[Author ORCID]{ G. Iacobello, https://orcid.org/0000-0002-0954-8545; L. Ridolfi, https://orcid.org/0000-0003-2947-8641; S. Scarsoglio, https://orcid.org/0000-0002-9427-6491}


\bibliographystyle{jfm}
\bibliography{Iacobello_etal_2021}

\end{document}